\documentclass[
superscriptaddress,twocolumn,
 amsmath,amssymb,
 aps,
prb,
]{revtex4-2}

\usepackage{dsfont}
\usepackage{graphicx}
\usepackage{dcolumn}
\usepackage{bm}
\usepackage{hyperref}
\hypersetup{
colorlinks=true,
	linkcolor=nrppurple,
	citecolor=nrppurple,
	urlcolor=nrppurple,
}
\usepackage{appendix}
\usepackage{fancyhdr}

\newcommand{\ket}[1]{\left| #1 \right>} 
\newcommand{\bra}[1]{\left< #1 \right|} 
\newcommand{\braket}[2]{\left< #1 | #2 \right>} 

\begin{document}
\definecolor{nrppurple}{RGB}{128,0,128}

	\title{Passive detection of Schwinger boson dynamics via a qubit}
\author{Ioannis Petrides}
\affiliation{Physical Sciences Division, College of Letters and Science, University of California, Los Angeles, CA 90095 USA}
\author{Arpit Arora}
\affiliation{Physical Sciences Division, College of Letters and Science, University of California, Los Angeles, CA 90095 USA}
\affiliation{Department of Electrical and Computer Engineering, UCLA, Los Angeles, CA, USA}
\author{Prineha Narang}
\affiliation{Physical Sciences Division, College of Letters and Science, University of California, Los Angeles, CA 90095 USA}
\affiliation{Department of Electrical and Computer Engineering, UCLA, Los Angeles, CA, USA}

	\date{\today}
	\begin{abstract}  
The quantum sensing landscape has been revolutionized by advanced technologies like superconducting circuits and qubit-based systems which have furthered the ability to probe and understand fundamental properties of quantum matter.
Here, we propose an integrated photonic device where a transmon qubit capacitively couples to a microwave cross-resonator, and the setup is employed for sensing of time reversal broken order in materials. 
In this sensing scheme, the transmon qubit plays a dual role as both a control element and a passive detector, while the photonic cross-resonator serves as the host for the sample, enabling a contact-free spectroscopic method suitable for studying materials where reliable electrical contacts are challenging to obtain, e.g., in van der Waal 2D heterostructures.
We show that by tuning the coupling strength and phase between the transmon and the cross-resonator, the system allows selective control over the interaction dynamics and leads to a highly sensitive detection method that can be compactly understood in terms of evolution of excited state population 
and 
quantum metric of the resonator-transmon hybrid state.
This architecture has the potential to host a wide range of quantum phenomena that can be precisely encoded in the dynamics of the transmon qubit and, in this way, 
potentially allows access to elusive aspects of correlated materials.
	\end{abstract}

\maketitle
\section{Introduction}

Microwave photonic platforms offer 
high precision 
sensing of quantum materials by 
light-matter hybrid states, which are tailored to encode the material’s unique properties in the dynamics of light~\cite{xia_high_2006,alegre_polarization-selective_2007,henderson_high-frequency_2008,hayashi_magneto-optical_2021,xia_detection_2015}. 
This has enabled the study of phenomena that are challenging using conventional methods, such as detecting weak magnetic dipoles~\cite{corr_kuiri_spontaneous_2022, corr_lee_theory_2019, corr_liu_tunable_2020}, spontaneous symmetry broken orders~\cite{ghosh2020recent}, pairing mechanisms in unconventional superconductors~\cite{Read.2000,kallin_chiral_2016,Poniatowski.2022a,Curtis.2022a,basov2011electrodynamics,han2025signatures,parra2025band}, or topological phases~\cite{petrides2022semiclassical,zhao2020axion,Curtis.2023a,Narang.2021}. 
Beyond quantum sensing, microwave photons are essential for qubit-based measurements, which underpin advanced quantum experiments and computational operations~\cite{nichol2017high,takahashi2013tomography,yu2022quantum}. 
In a typical set up, a transmon qubit can be coupled to a resonator, such as a microwave cavity, allowing interactions to be probed with minimal disturbance to the system~\cite{hofheinz_generation_2008,devoret2005implementing,valles2023trimming,soltani2017efficient,fan2018superconducting}. 
By integrating the microwave photon-based interactions with qubit-based measurements can further enhance the precision standards, where the qubit effectively provides a control knob over the quantum process tomography of photon mode dynamics.

Here, we propose a device that integrates a transmon qubit with a photonic cross-resonator, and demonstrate how qubit control of photon dynamics enables sensing of time reversal broken order in quantum materials. 
The cross-resonator setup, recently proposed for detecting time-reversal symmetry breaking~\cite{petrides2025probing}, is able to isolate individual components of a material's complex refractive index while minimizing Cram\'er-Rao bound on parameter estimation variance. 
This non-invasive approach offers a significant upper hand over conventional methods, which often face limitation such as degradation of sample or the requirement of high quality contacts with low resistance~\cite{wang2013one,cao_unconventional_2018,sajadi_gate-induced_2018}. 
Meanwhile, qubit-based techniques
offer high precision control over measurement protocols~\cite{vallabhapurapu2023high,childress2013diamond,wang2020electrical,oliver2005mach,khabiboulline2019optical}. 
Leveraging the advantages of these two platforms through controlled interactions provides a powerful probing framework for exploring the electromagnetic properties of quantum materials. 
Quantum geometry plays a central role in this process, capturing how the material’s dielectric response are encoded within the system’s dynamics~\cite{carinena2015geometry,ashtekar1999geometrical}. 
By mapping these changes to the qubit’s evolution allows it to serve as a highly sensitive detector, capable of precision decoding of the induced dynamics and with the ability to maximize the information that can be extracted.

Our setup consists of a transmon qubit capacitively coupled to a planar photonic cross-resonator which serves as the host for a material sample. 
The transmon qubit plays a dual role as both a control element and a passive detector, with its energy levels controlled by external parameters, i.e., the magnetic flux and radio-frequency (RF) signals. 
The photonic modes of the cross-resonator interact with the sample through its dielectric response, modifying the photon dynamics based on the sample's complex refractive index. 
By tuning the coupling strength and phase between the transmon and the cross-resonator, the system allows access to the interaction dynamics, described in terms of a ''coupling vector” that, when aligned with the cross-resonator’s coupling vector, amplifies the sensitivity to the dielectric and magnetic properties of the sample. 
In this regime the system hosts an uncoupled bright state which hybridizes with other states when perturbations are such as broken time reversal order are present. Our approach quantifies the symmetry breaking with evolution of population and quantum geometry of resonator-transmon hybrid state. 
Finally, the transmon’s evolution then serves as a passive readout mechanism, enabling a contact-free spectroscopic technique.

\section{Device Description}

\begin{figure}[tb]
    \centering
    \includegraphics[width= 0.5\textwidth]{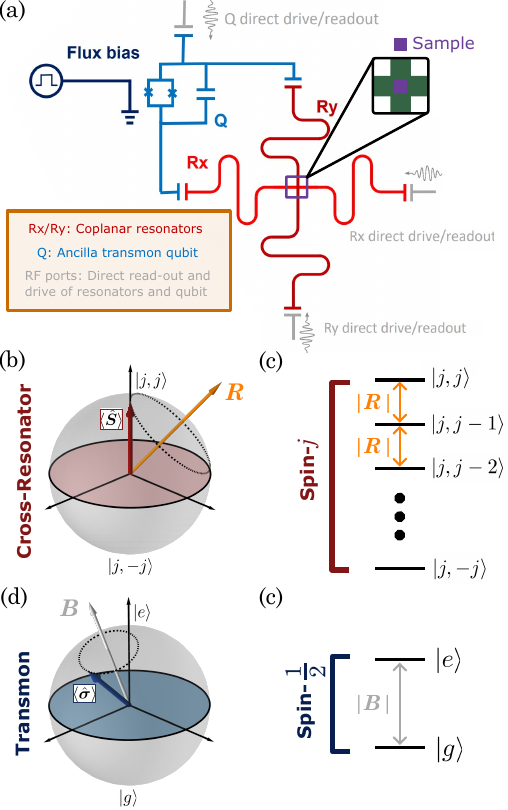}
    \caption{
    Proposed cross resonator-transmon device:
    (a) A transmon qubit (blue accents) capacitively coupled to a cross-resonator device (red accents), where Rx and Ry denote the two resonators.
    The transmon's energy splitting can be controlled via a magnetic flux threading. 
    The device is externally driven and measured via capacitive coupling to transmission lines (gray). 
    (b) The evolution of the transmon's (cross-resonator's) state vector $\langle\hat{\bm \sigma}\rangle$ ($\langle\hat{\bm S}\rangle$) is equivalent to the precession around the rotation vector $\bm B$ ($\bm R$).}
    \label{fig:fig1}
\end{figure} 
We consider a structured device made by a cross-resonator system in a planar geometry (as previously described in Ref.~\cite{petrides2025probing}) coupled to a transmon qubit via capacitive contacts, see Fig.~\ref{fig:fig1}(a).
We assume that the photonic modes in the cross-resonator interact via a small dielectric sample of interest, placed at the intersection, see inset of Fig.~\ref{fig:fig1}(a).
The sample is well described by a distribution, hereafter denoted by $f(\bm r)$, susceptibility tensor $\chi_{ij} $, and a conductivity tensor $\sigma_{ij} = \varepsilon^{ij}\sigma_H$, where $\sigma_H$ is the Hall conductivity and $\varepsilon^{ij}$ is the Levi-Civita tensor. 
Note that $\sigma_H\neq 0 $ when time reversal symmetry is broken.
To maintain generality, we set the diagonal conductivity to zero, integrating its effect into the device’s coherence time. 
Additionally, we assume that each resonator sustains a single mode, with the electric field extending into free space to enable evanescent coupling with the sample.

The cross-resonator is described by the Hamiltonian operator~\cite{petrides2025probing}
\begin{equation}
  \hat{H}_{C} =\omega_0 \hat{S}_0+\bm R \cdot \hat{\bm S}
\end{equation}
where $\omega_0$ is the central frequency, and $\bm R$ is a vector determined by dielectric and magnetic response of the sample of interest.
Specifically, both the $\bm R_x$ and $\bm R_z$ components are directly proportional to the real part of the complex refractive index that characterizes the sample, namely, the real susceptibility, as well as to any mode splitting and geometrical coupling due to imperfections of the cross-resonator device.
On the other hand, a nonzero $\bm R_y$ component arises only when the sample inherently breaks time-reversal symmetry, as in the case of $\sigma_H\neq 0$. 
Hereafter, we refer to $\bm R$ as the cross-resonator's "rotation vector" which maps the dielectric and magnetic response of the sample onto a Bloch spehere. 
Without loss of generality, we represent it as $\bm R={|\bm R|}\left(\cos(\theta_{0})\sin(\phi_{0}),\sin(\theta_{0})\sin(\phi_{0}),\cos(\phi_{0})\right)$, where $|\bm R|$ is the magnitude, and $\theta_{0}$ ($\phi_{0}$) is the polar (azimuthial) angle.
The spin operators $\hat{\bm S}$ corresponding to the Bloch sphere maping $\bm R$ are given by
\begin{eqnarray}
\begin{array}{c}
    \hat{ S}_x =\frac{1}{2}\left(\hat{ a}^\dagger\hat{ b}^{\empty}+h.c.\right)\,,      \hat{ S}_y =\frac{i}{2}\left(\hat{ a}^\dagger\hat{ b}^{\empty}- h.c.\right)\,,\\ \text{and}\,
      \hat{ S}_z =\frac{1}{2}\left(\hat{ a}^\dagger\hat{ a}^{\empty}-\hat{ b}^\dagger \hat{ b}^{\empty} \right)
\end{array}
\end{eqnarray} 
with commutation relations $\left[\hat S_i , \hat S_j\right] = i \epsilon^{ijk}\hat S_k$ and $\hat { S}_0 = \hat{a}^\dagger \hat{a}+ \hat{b}^\dagger\hat{b}$ is the total number of photons. 
For a fixed number of photons $n =\langle\hat{S}_0\rangle$ in the cross-resonator device, the eigenstates form a spin-$j$ representation given by $|j,m\rangle$, where $j=n/2$ is the total pseudo-spin and $m=\langle \hat{S}_z\rangle$ is the associated z-component, see Figs.~\ref{fig:fig1}(b) and (c).

The transmon qubit is described by the Hamiltonian
\begin{eqnarray}
    \hat{H}_{T} = \bm B \cdot \hat{\bm \sigma}
\end{eqnarray}
where $\bm B=(\text{Re}\,\Omega, \text{Im}\,\Omega, \Phi)$ is a three-dimensional vector, hereafter refer to as the transmon's "rotation vector", determined by the flux bias $\Phi$ and radio-frequency (RF) driving $\Omega$ of the transmon, while
\begin{eqnarray}
\begin{array}{c}
    \hat{ \sigma}^+ =\ket{e}\bra{g}\,,  \hat{ \sigma}^- =\ket{g}\bra{e}\,, \text{and}\,
      \hat{\sigma}_z = \ket{e}\bra{e}-\ket{g}\bra{g}
\end{array}
\end{eqnarray} 
are the spin-$\frac{1}{2}$ representations of the SU(2) algebra. 
The transmon possess two degrees of freedom where its corresponding state vector can be represented on a 2-sphere, see Fig.~\ref{fig:fig1}(d) and (e). 
The ensued dynamics of both the transmonic and photonic degrees of freedom are equivalent to spin precession around an effective magnetic field, where the spin is represented by the state vector and the magnetic field by the corresponding rotation vector.

\begin{figure}[tb]
    \centering
    \includegraphics[width= 0.5\textwidth]{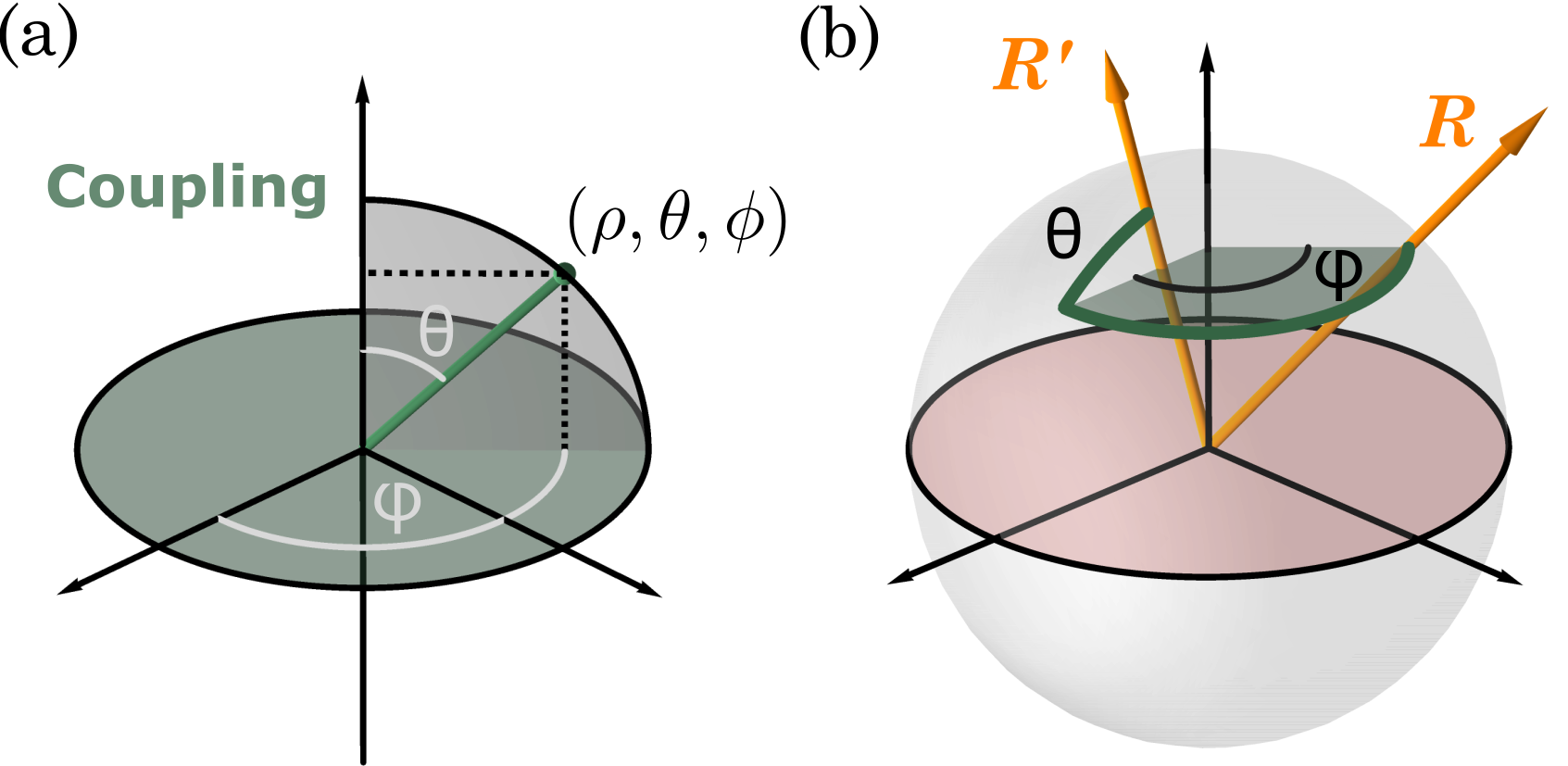}
    \caption{
    (a) The coupling between the transmon and cross-resonator is compactly represented by a vector in a three-dimensional sphere with radius $g=\sqrt(g_a^2 + g_b ^2)/2$ given by the root mean squared interaction strength, polar angle $\theta = \arctan\frac{|g_b|}{|g_a|}$ and azimuthial angle $\phi = \frac{1}{2}\arg (g_a g^*_b)$.
    (b) Partially diagonalizing the coupling Hamiltonian $\hat{H}_{coupling}$ results in the rotation of the cross-resonator's rotation vector $\bm R$ according to the coupling strengths $g_a$ and $g_b$, and the relative phase difference $\phi$.}
    \label{fig:fig2}
\end{figure}

The transmon serves both as a passive read-out system, as well as a control knob for the dynamics via its coupling to the cross-resonator. 
The latter permits the exchange of excitations between the photonic modes and the transmon with a tunable strength and phase, with the coupling Hamiltonian given by 
\begin{equation}
    \hat{H}_{coupling} = \left(g_a e^{i \frac{\phi}{2}}\hat{a}^\dagger+g_b e^{-i \frac{\phi}{2}}\hat{b}^\dagger\right)\sigma^- + \text{h.c}
\end{equation}
where $g_a$ and $g_b$ are the coupling strengths to each photonic mode and $\phi$ is the relative phase difference.
Physically, the coupling Hamiltonian $\hat{H}_{coupling}$ describes the excitation of the photonic modes with a simultaneous de-excitation of the transmon, and vice-versa.
The coupling between the transmon and cross-resonator device can be represented geometrically by a vector on a three-dimensional ball, hereafter dubbed the "coupling vector", where the polar and azimuthial angles are given by $\theta = \arctan\frac{|g_b|}{|g_a|}$ and the relative phase $\phi$, respectively; the radius is determined by the root squared interaction strength $\rho=\sqrt(g_a^2 + g_b ^2)$, see Fig~\ref{fig:fig2}(a).

\section{Spectral analysis }

To demonstrate important features of the system 
we partially diagonalize the coupling Hamiltonian $\hat{H}_{coupling}$ via the transformation
\begin{eqnarray}
\label{eq:transformation}
    \hat{T} = e^{i \frac{\phi}{2}\hat{S}_z}e^{i \frac{\theta}{2}\hat{S}_y}
\end{eqnarray}
where $\theta $ and $\phi$ are determined by the coupling vector. 
The transformation in Eq.~(\ref{eq:transformation}) introduces a rotation of the cross-resonator's rotation vector along the azimuthial and polar directions, with the cross-resonator's new rotation vector given by $ \bm R{'}  = \text{Rot}_{\theta,\phi}\left(\bm R\right) $, where $\text{Rot}_{\theta,\phi}$ is the rotation matrix in three dimensions, see Fig~\ref{fig:fig2}(b).
In this basis, the new bosonic operators given by $\hat{A}^\dagger =\rho^{-1}\left(g_ae^{i \frac{\phi}{2}} \hat{a}^\dagger+g_b e^{-i \frac{\phi}{2}}\hat{b}^\dagger\right)$ and $\hat{B}^\dagger =\rho^{-1}\left(-g_b e^{i \frac{\phi}{2}} \hat{a}^\dagger+g_a e^{-i \frac{\phi}{2}}\hat{b}^\dagger\right)$, respectively.
Importantly, the coupling Hamiltonian takes the form 
\begin{eqnarray}
    \hat{H}'_{coupling} = \rho \hat{A}^\dagger\sigma^- + \text{h.c.}\,,
\end{eqnarray}
as one of the bosonic degrees of freedom is decoupled from the transmon reducing the number of interacting terms and making it easier to analyze the system's behavior.

Our passive detection scheme is based on initially exciting the transmon and subsequently measuring its dynamics. 
Such dynamics are confined in a three dimensional submanifold of product states given by {${\{|00e\rangle,|10g\rangle,|01g\rangle\}}$}, with the corresponding Hamiltonian representation in the partially diagonalized basis given by
\begin{eqnarray}
  \langle T\hat{H}T^{-1}\rangle =  \left( \begin{array}{cc}
         \frac{\bm B_z}{2}&\begin{array}{c} g \hspace{35pt} 0\end{array} \\
         \begin{array}{c} g \\ 0\end{array} &\left(\omega_0 -\frac{\bm B_z}{2}\right)\mathds{1}_{2\times2} +\bm R'\cdot \bm \sigma 
    \end{array}\right)
    \label{eq:HamRepr}
\end{eqnarray}
where $\hat{H}=\hat{H}_C + \hat{H}_T +\hat{H}_{coupling}$ is the total Hamiltonian, $\bm \sigma =\left(\sigma_x,\sigma_y,\sigma_z\right)$ are the three Pauli matrices, and $\mathds{1}_{2\times2}$ is the $2\times2$ identity matrix.
In obtaining Eq.~\ref{eq:HamRepr}, we have assumed, without loss of generality, that the qubit is not driven, i.e.,  $\Omega=0$, and energy levels are completely determined by $\Phi$, i.e., $\textbf{\textit{B}} = (0,0,\textbf{\textit{B}}_z = \Phi)$.

\begin{figure}[tb]
    \centering
    \includegraphics[width= 0.5\textwidth]{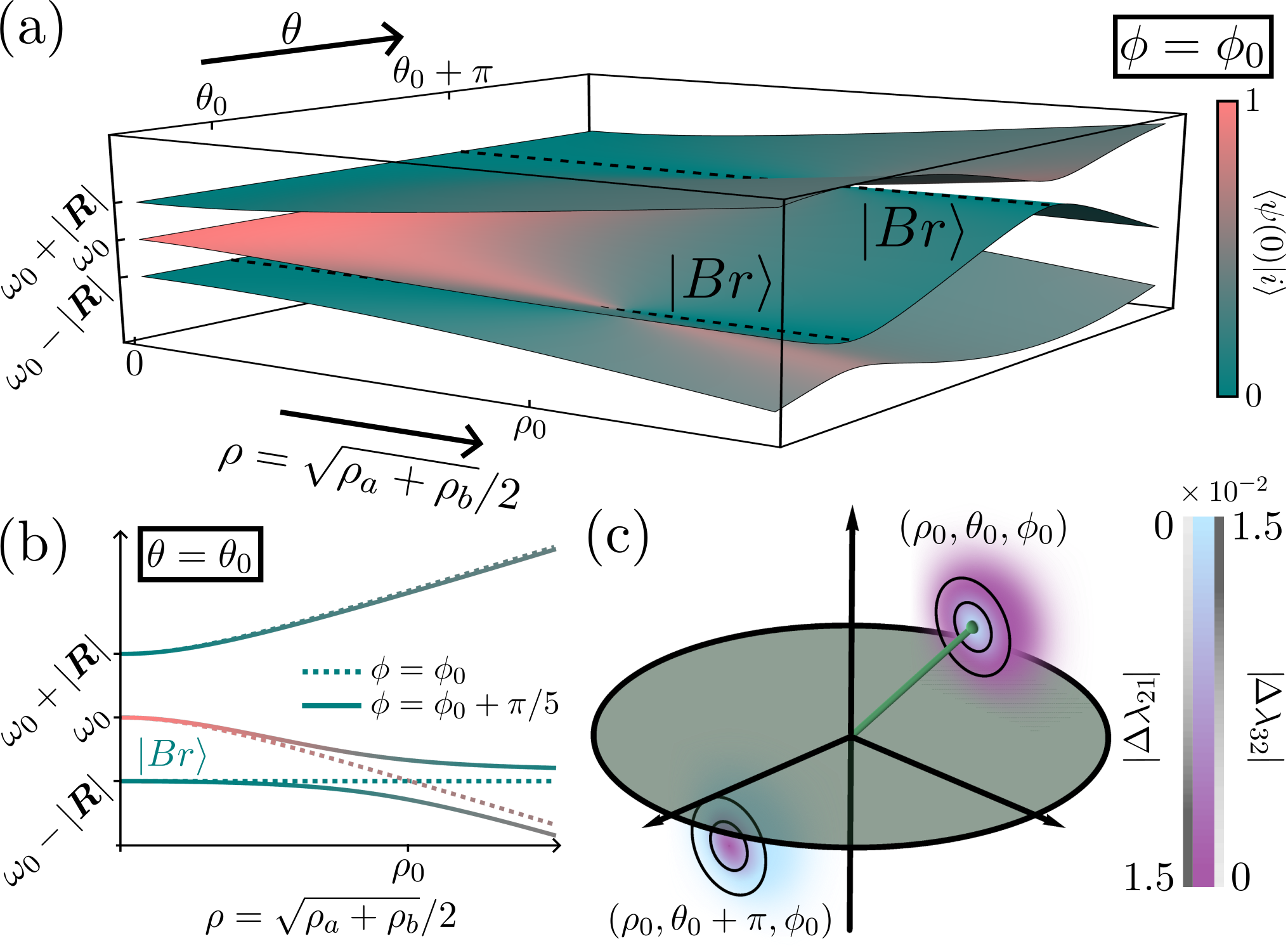}
    \caption{
    (a) The spectrum of the three-dimensional subspace spanned by initially exciting the transmon, and where the coupling vector is fixed to $\phi-\phi_0$.
    (b) Similarly, the spectrum of the three-dimensional subspace for $\theta=\theta_0$ for two differnet values of $\phi=\phi_0$ or $\phi_0+\pi/5$.
    (c) Density plot of the energy difference between the eigenenergies shows a degeneracy when the coupling vector has the same coordinates as the cross-resonator's rotational vector.   
    }
    \label{fig:fig3}
\end{figure}

To proceed, we consider the central frequency of cross-resonator device in resonance with the transmon such that $\bm B_z = \omega_0$.
The transmon-cross resonator structured device described by Hamiltonian in Eq.~\ref{eq:HamRepr} is initialized in the state $|\psi(0)\rangle = |00e\rangle$. 
The energy spectrum of the system as a function of the coupling parameters, $\rho$ and $\theta$, for fixed $\phi = \phi_0$ is shown in Fig.~\ref{fig:fig3}(a) and (b). 
Without any coupling, i.e., $\rho=0$, the spectrum has three separate eigenenergies at $\omega_0$ and $\omega_0\pm |\bm R|$.
Turning the coupling on results in the mixing of the photonic and transmonic degrees of freedom according to the strength $\rho$ and direction $(\theta,\phi)$ of the coupling vector.
Importantly, when the coupling vector aligns with the cross-resonator's rotation vector, i.e., $\phi=\phi_0$ and $\theta=\theta_0 +\nu \pi\, \text{mod}\, 2\pi$, where $\nu$ is an integer, one of the photonic degrees of freedom decouples, resulting in a single bright state denoted as $\ket{Br}$.
Interestingly, when the coupling vector $(\rho,\theta,\phi)$ exactly matches the cross-resonator's rotation vector $\bm R$ in length and direction, either in the parallel ($\theta=\theta_0$) or anti-parallel ($\theta=\theta_0 +\pi$) configuration, two of the energies become degenerate, see Fig.~\ref{fig:fig3}(c). Additionally, given that Eq.~\ref{eq:HamRepr} diagonalizes to give the dressed states $|i\rangle$ with energy $\lambda_i$, we calculate $\langle\psi(0)|i\rangle$, to track the weight of initial state in the dressed states as photon and transmon degrees of freedom are coupled. 
As shown in Fig.~\ref{fig:fig3}(a) and (b), $c_i = \langle \psi(0) | i\rangle$ is concentrated at energy $\omega_0$ for $\rho = 0$,i.e., the state is fully transmonic, and partitions to other states only as the coupling is turned on.

\section{Dynamics and sensing of time reversal symmetry breaking }

To unwrap the information carried by the system we study the dynamical properties of the device. 
The evolution of initial state $|\psi\rangle$
is determined by 
\begin{eqnarray}
    \ket{\psi(t)} = \hat{U}(t)\ket{\psi(0)} = \sum_{i} c_i e^{it\lambda_i}\ket{i}
\end{eqnarray}
where $\hat{U}(t)=e^{it\hat{H}}$ is the evolution operator, and as defined above $\ket{i}$ are the dressed states with corresponding energies $\lambda_i$, and $c_i = \langle \psi(0)|i\rangle$ are coefficients determined by the initial state. 
The evolution of the population in the excited state, determined by 
$\rho_e(t) = |\braket{\psi(t)}{\psi(0)}|^2$,
is given by 

\begin{eqnarray}
    \begin{array}{rl}
       \rho_e(t)   &= \sum\limits_{i,j} e^{i \Delta\lambda_{ij}t}|c_i|^2 |c_j|^2\\
        &=1-\sum\limits_{i>j}4|c_i|^2|c_j|^2\sin^2\left(\frac{\Delta\lambda_{ij}}{2}t\right) 
    \end{array}
\end{eqnarray}
where $\Delta\lambda_{ij}=\lambda_i-\lambda_j$ is the energy difference between the dressed states.
The transmon's excited state follows a sinusoidal evolution with amplitude and frequency determined by both the coupling and the direction of the cross-resonator's rotation vector. This behavior demonstartes the coherent exchange of energy between the transmon and cross cavity system. 
The periodic evolution of the transmon's excited state is a key feature of the underlying passive detection scheme. 
By measuring the amplitude and frequency of these oscillations, one can infer information about the coupling strength and the sample's dielectric properties.

\begin{figure}[t!]
    \centering
    \includegraphics[width= 0.5\textwidth]{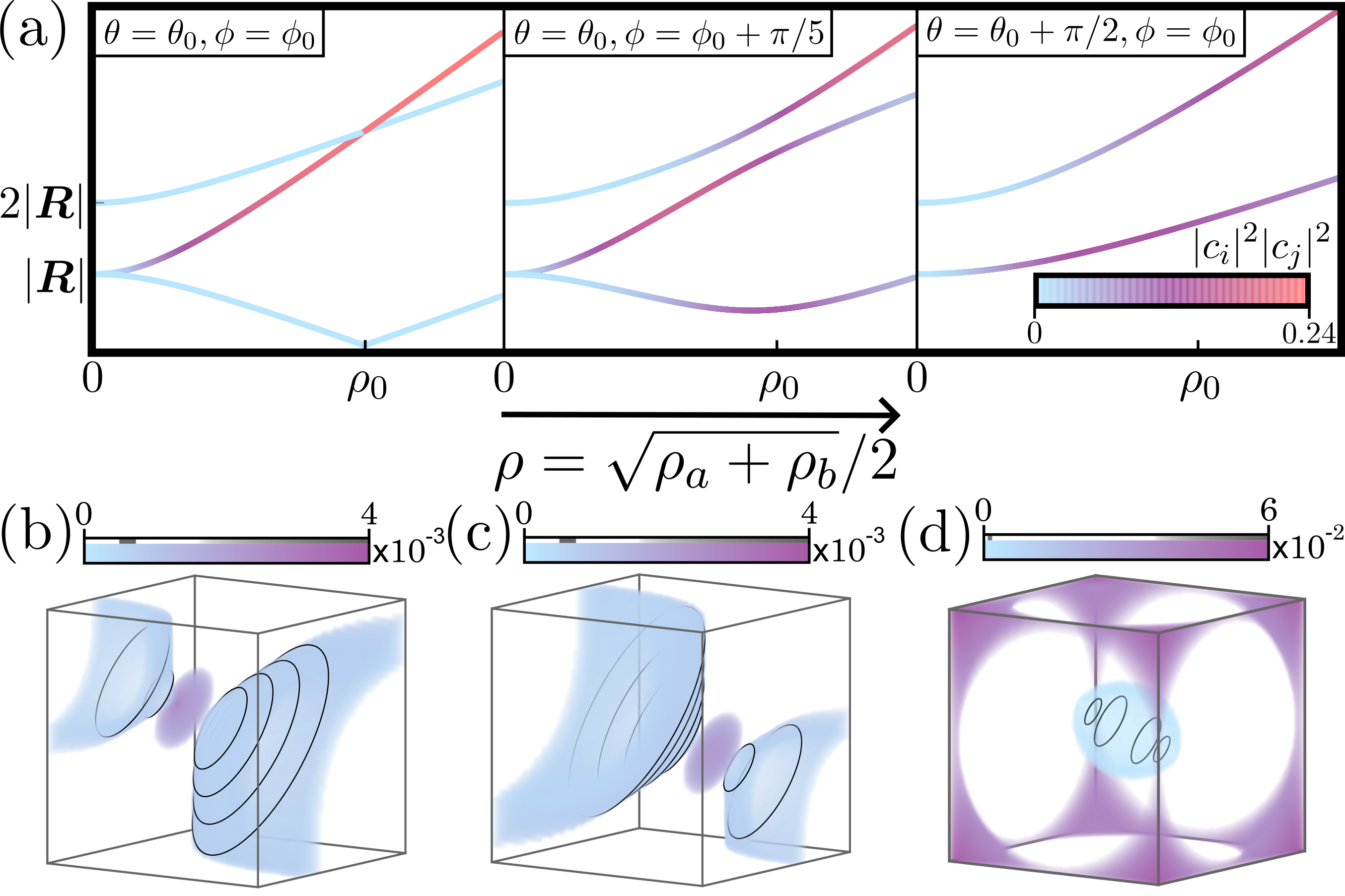}
    \caption{ (a) The Fourier transform of the transmon's dynamics. 
    Depending on the coupling vector, the transmon coupled to the photonic modes and its population oscillates with a maximum of three different frequencies.
    The amplitude of the oscillation associated to each frequency depends on the overlap of the initial state with the systems hybrid modes.
    (b) The quantum metric extracted from the frequency and amplitude of oscillations. 
    For each pair of eigenmodes, the quantum metric is maximized depending on the direction and magnitude of the cross-resonator's rotation vector.
    }
    \label{fig:fig4}
\end{figure}

For the chosen initial state, the transmon's population $\rho_e(t)$ is decomposed into a collection of peaks in frequency space, each centered around the energy difference $\Delta \lambda_{ij}$ with corresponding amplitude determined by $|c_i|^2|c_j|^2$
\begin{eqnarray}
    FT\left[\rho_e(t)\right] = \sqrt{2\pi}\sum\limits_{i, j}|c_i|^2|c_j|^2\delta(\omega-\Delta\lambda_{ij})\,.
\end{eqnarray}
When the coupling vector's direction matches the cross-resonator's rotation vector, i.e., $\theta = \theta_{0}$ and $\phi=\phi_{0}$, there is only one accessible state as the bright state $|Br\rangle$ is decoupled from the transmon, see Fig.~\ref{fig:fig4}(a) \textit{left}.
One the other hand, when the coupling vector's direction is perpendicular to the cross-resonator's rotation vector, we find three maximally separated energies, hence, two of the Fourier peaks will coincide, see Fig.~\ref{fig:fig4}(a) \textit{right}. 
For arbitrary coupling, the frequency and amplitude of the Fourier spectrum depends on the total strength $\rho$ and phases $\theta$ and $\phi$, see Fig.~\ref{fig:fig4}(a) \textit{middle}. 
These features can be utilized to operate this device, for instance, as a sensitive probe for time reversal broken order.
We benchmark the device with the signal shown in Fig.~\ref{fig:fig4} (a) where time reversal breaking and other dielectric properties can be tracked by the coupling of $|Br\rangle$ with other states, e.g., Fig.~\ref{fig:fig4} (b).

To comment on the sensitivity and optimal operation regime, we calculate the quantum metric
\begin{eqnarray}
\begin{array}{rl}
    Q& = \braket{\partial_t \psi(t)}{\partial_t \psi(t)}-|\braket{\partial_t \psi(t)}{\psi(t)}|^2
\end{array}
\end{eqnarray}
which is related to the quantum Fisher information. 
The quantum metric quantifies how sensitive the system's state $|\psi(t)\rangle$ is to changes in time, which indirectly reflects its sensitivity to changes in the system's parameters (e.g., coupling strength, phase, or cross-cavity's rotation vector). 
We track the dynamical evolution of the transmon and optimize the quantum geometry of the resulting hybrid states using the parameters of the coupling vector, specifically, we use the associated quantum metric (see Appendix~\ref{appx:QM}) 
\begin{eqnarray}
\begin{array}{rl}
    Q& =\sum\limits_{i>j}g_{ij}(\rho,\theta,\phi)
\end{array}
\end{eqnarray}
where $g_{ij} (\rho,\theta,\phi) = \Delta\lambda_{ij}^2|c_i|^2|c_j|^2$, and $i$ and $j$ denote the set of dressed states.
The dependence of the quantum metric $g_{ij}$ on the coupling vector's strength and direction $(g,\theta,\phi)$, shown in Fig.~\ref{fig:fig4}(b)-(d), has maxima/minima when the coupling vector's direction matches the cross-resonator's rotation vector, i.e., when $\theta = \theta_{0}$ and $\phi=\phi_{0}$. 
Such maxima/minima become divergent when the coupling strength is equal to $|\bm R|$, i.e., when $\rho=\rho_0$. 
This consistent with the sensitive regime pointed out in terms of population of excited state, c.f. Fig.~\ref{fig:fig4}(a).

\section{Conclusions}
Here we propose an integrated photonic device composed by a transmon and a cross-resonator designed to enable passive detection of quantum material properties, e.g., time reversal broken order. 
The cross-resonator 
houses the sample of interest that induces a coupling of the photonic modes according to its complex dielectric and magnetic response.
The transmon is used both as a drive, by initializing the system in its excited state, as well as passive detector that monitors the dynamics of the dressed states.
By selectively controlling the coupling between the transmon and the cross-resonator, we show that the 
time reversal breaking of the sample 
is encoded in the evolution of the transmon population number. We systematically obtain the optimal regime of operation where when the coupling vector aligns with $\bm R$, the system hosts an uncoupled bright state. 
Tracking the hybridization of the bright mode in presence of perturbations serves as a demonstration for operation of the device.
This serves as a passive detector which is capable of discerning the time reversal broken order and complex refractive index via a destructive/constructive interference of the system dynamics.

Looking ahead, this architecture contributes to the toolkit of new microwave photonics device designs, particularly for interacting with correlated matter~\cite{phan2022detecting,bottcher2024circuit,tanaka2024kinetic,banerjee2024superfluid,kreidel2024measuring,arora2025chiral,arora2025quantum,dirnegger2025nonlinear,di2024infrared,jin2025exploring,thiemann2018single}. Levergaing qubit dynamics at its core, the proposed platform also opens avenues for investigating nonequilibrium systems and phase transitions. 
Additionally, the integration of advanced control techniques, such as machine learning algorithms for parameter optimization, could further enhance the precision and versatility of the device.

\appendix
\section{Quantum metric}\label{appx:QM}

The quantum metric associated to the state $\ket{\psi(t)}= \sum_{i} c_i e^{it\lambda_i}\ket{i}$ is defined as
\begin{eqnarray}
\begin{array}{rl}
    Q& = \braket{\partial_t \psi(t)}{\partial_t \psi(t)}-|\braket{\partial_t \psi(t)}{\psi(t)}|^2
\end{array}
\end{eqnarray}
After partial derivation with respect to the time coordinate, the first term is given by
\begin{eqnarray}
    \braket{\partial_t \psi(t)}{\partial_t \psi(t)} = \sum\limits_i \lambda_i^2|c_i|^2
\end{eqnarray}
and the second term by
\begin{eqnarray}
    |\braket{\partial_t\psi(t)}{ \psi(t)} |^2= \sum\limits_{i,j} \lambda_i\lambda_j|c_i|^2|c_j|^2
\end{eqnarray}
Combining the above two results in 
\begin{eqnarray}
    \begin{array}{rl}
        Q & = \sum\limits_i \lambda_i^2 |c_i|^2 (1-|c_i|^2) - \sum\limits_{i\ne j}\lambda_i\lambda_j |c_i|^2|c_j|^2\\
         & = \sum\limits_{i\ne j} \lambda_i ^2 |c_i|^2 |c_j|^2 - 2\sum\limits_{i> j}\lambda_i\lambda_j |c_i|^2|c_j|^2\\
         & = \sum\limits_{i> j}\left(\lambda_i^2+\lambda_j^2 - 2\lambda_i\lambda_j\right)|c_i|^2|c_j|^2\\
         &=\sum\limits_{i> j}\left(\lambda_i-\lambda_j \right)^2|c_i|^2|c_j|^2
    \end{array}
\end{eqnarray}
where we have used the normalization condition $\sum_i |c_i|^2=1$.
The last equality can be rewritten as $Q=\sum\limits_{i>j}g_{ij}(\rho,\theta,\phi)$ by defining \begin{eqnarray}
    g_{ij} (\rho,\theta,\phi) = \Delta\lambda_{ij}^2|c_i|^2|c_j|^2
\end{eqnarray}

\begin{acknowledgments}
\textbf{Acknowledgments}
This work is supported by the Quantum Science Center (QSC), a National Quantum Information Science Research Center of the U.S. Department of Energy (DOE). 
P.N. gratefully acknowledges support from the Gordon and Betty Moore Foundation grant No. 8048 and from the John Simon Guggenheim Memorial Foundation (Guggenheim Fellowship). 
\end{acknowledgments}

\bibliography{ref}

\begin{thebibliography}{50}%
\makeatletter
\providecommand \@ifxundefined [1]{%
 \@ifx{#1\undefined}
}%
\providecommand \@ifnum [1]{%
 \ifnum #1\expandafter \@firstoftwo
 \else \expandafter \@secondoftwo
 \fi
}%
\providecommand \@ifx [1]{%
 \ifx #1\expandafter \@firstoftwo
 \else \expandafter \@secondoftwo
 \fi
}%
\providecommand \natexlab [1]{#1}%
\providecommand \enquote  [1]{``#1''}%
\providecommand \bibnamefont  [1]{#1}%
\providecommand \bibfnamefont [1]{#1}%
\providecommand \citenamefont [1]{#1}%
\providecommand \href@noop [0]{\@secondoftwo}%
\providecommand \href [0]{\begingroup \@sanitize@url \@href}%
\providecommand \@href[1]{\@@startlink{#1}\@@href}%
\providecommand \@@href[1]{\endgroup#1\@@endlink}%
\providecommand \@sanitize@url [0]{\catcode `\\12\catcode `\$12\catcode `\&12\catcode `\#12\catcode `\^12\catcode `\_12\catcode `\%12\relax}%
\providecommand \@@startlink[1]{}%
\providecommand \@@endlink[0]{}%
\providecommand \url  [0]{\begingroup\@sanitize@url \@url }%
\providecommand \@url [1]{\endgroup\@href {#1}{\urlprefix }}%
\providecommand \urlprefix  [0]{URL }%
\providecommand \Eprint [0]{\href }%
\providecommand \doibase [0]{https://doi.org/}%
\providecommand \selectlanguage [0]{\@gobble}%
\providecommand \bibinfo  [0]{\@secondoftwo}%
\providecommand \bibfield  [0]{\@secondoftwo}%
\providecommand \translation [1]{[#1]}%
\providecommand \BibitemOpen [0]{}%
\providecommand \bibitemStop [0]{}%
\providecommand \bibitemNoStop [0]{.\EOS\space}%
\providecommand \EOS [0]{\spacefactor3000\relax}%
\providecommand \BibitemShut  [1]{\csname bibitem#1\endcsname}%
\let\auto@bib@innerbib\@empty
\bibitem [{\citenamefont {Xia}\ \emph {et~al.}(2006)\citenamefont {Xia}, \citenamefont {Maeno}, \citenamefont {Beyersdorf}, \citenamefont {Fejer},\ and\ \citenamefont {Kapitulnik}}]{xia_high_2006}%
  \BibitemOpen
  \bibfield  {author} {\bibinfo {author} {\bibfnamefont {J.}~\bibnamefont {Xia}}, \bibinfo {author} {\bibfnamefont {Y.}~\bibnamefont {Maeno}}, \bibinfo {author} {\bibfnamefont {P.~T.}\ \bibnamefont {Beyersdorf}}, \bibinfo {author} {\bibfnamefont {M.~M.}\ \bibnamefont {Fejer}},\ and\ \bibinfo {author} {\bibfnamefont {A.}~\bibnamefont {Kapitulnik}},\ }\bibfield  {title} {\bibinfo {title} {High {Resolution} {Polar} {Kerr} {Effect} {Measurements} of {Sr} 2{RuO}4 : {Evidence} for {Broken} {Time}-{Reversal} {Symmetry} in the {Superconducting} {State}},\ }\href@noop {} {\bibfield  {journal} {\bibinfo  {journal} {Physical Review Letters}\ }\textbf {\bibinfo {volume} {97}},\ \bibinfo {pages} {167002} (\bibinfo {year} {2006})}\BibitemShut {NoStop}%
\bibitem [{\citenamefont {Alegre}\ \emph {et~al.}(2007)\citenamefont {Alegre}, \citenamefont {Santori}, \citenamefont {Medeiros-Ribeiro},\ and\ \citenamefont {Beausoleil}}]{alegre_polarization-selective_2007}%
  \BibitemOpen
  \bibfield  {author} {\bibinfo {author} {\bibfnamefont {T.~P.~M.}\ \bibnamefont {Alegre}}, \bibinfo {author} {\bibfnamefont {C.}~\bibnamefont {Santori}}, \bibinfo {author} {\bibfnamefont {G.}~\bibnamefont {Medeiros-Ribeiro}},\ and\ \bibinfo {author} {\bibfnamefont {R.~G.}\ \bibnamefont {Beausoleil}},\ }\bibfield  {title} {\bibinfo {title} {Polarization-selective excitation of nitrogen vacancy centers in diamond},\ }\href {https://doi.org/10.1103/PhysRevB.76.165205} {\bibfield  {journal} {\bibinfo  {journal} {Physical Review B}\ }\textbf {\bibinfo {volume} {76}},\ \bibinfo {pages} {165205} (\bibinfo {year} {2007})}\BibitemShut {NoStop}%
\bibitem [{\citenamefont {Henderson}\ \emph {et~al.}(2008)\citenamefont {Henderson}, \citenamefont {Ramsey}, \citenamefont {Quddusi},\ and\ \citenamefont {del Barco}}]{henderson_high-frequency_2008}%
  \BibitemOpen
  \bibfield  {author} {\bibinfo {author} {\bibfnamefont {J.~J.}\ \bibnamefont {Henderson}}, \bibinfo {author} {\bibfnamefont {C.~M.}\ \bibnamefont {Ramsey}}, \bibinfo {author} {\bibfnamefont {H.~M.}\ \bibnamefont {Quddusi}},\ and\ \bibinfo {author} {\bibfnamefont {E.}~\bibnamefont {del Barco}},\ }\bibfield  {title} {\bibinfo {title} {High-frequency microstrip cross resonators for circular polarization electron paramagnetic resonance spectroscopy},\ }\href {https://doi.org/10.1063/1.2957621} {\bibfield  {journal} {\bibinfo  {journal} {Review of Scientific Instruments}\ }\textbf {\bibinfo {volume} {79}},\ \bibinfo {pages} {074704} (\bibinfo {year} {2008})}\BibitemShut {NoStop}%
\bibitem [{\citenamefont {Hayashi}\ \emph {et~al.}(2021)\citenamefont {Hayashi}, \citenamefont {Okamura}, \citenamefont {Kanazawa}, \citenamefont {Yu}, \citenamefont {Koretsune}, \citenamefont {Arita}, \citenamefont {Tsukazaki}, \citenamefont {Ichikawa}, \citenamefont {Kawasaki}, \citenamefont {Tokura},\ and\ \citenamefont {Takahashi}}]{hayashi_magneto-optical_2021}%
  \BibitemOpen
  \bibfield  {author} {\bibinfo {author} {\bibfnamefont {Y.}~\bibnamefont {Hayashi}}, \bibinfo {author} {\bibfnamefont {Y.}~\bibnamefont {Okamura}}, \bibinfo {author} {\bibfnamefont {N.}~\bibnamefont {Kanazawa}}, \bibinfo {author} {\bibfnamefont {T.}~\bibnamefont {Yu}}, \bibinfo {author} {\bibfnamefont {T.}~\bibnamefont {Koretsune}}, \bibinfo {author} {\bibfnamefont {R.}~\bibnamefont {Arita}}, \bibinfo {author} {\bibfnamefont {A.}~\bibnamefont {Tsukazaki}}, \bibinfo {author} {\bibfnamefont {M.}~\bibnamefont {Ichikawa}}, \bibinfo {author} {\bibfnamefont {M.}~\bibnamefont {Kawasaki}}, \bibinfo {author} {\bibfnamefont {Y.}~\bibnamefont {Tokura}},\ and\ \bibinfo {author} {\bibfnamefont {Y.}~\bibnamefont {Takahashi}},\ }\bibfield  {title} {\bibinfo {title} {Magneto-optical spectroscopy on {Weyl} nodes for anomalous and topological {Hall} effects in chiral {MnGe}},\ }\href {https://doi.org/10.1038/s41467-021-25276-1} {\bibfield  {journal} {\bibinfo  {journal} {Nature Communications}\ }\textbf {\bibinfo {volume}
  {12}},\ \bibinfo {pages} {5974} (\bibinfo {year} {2021})}\BibitemShut {NoStop}%
\bibitem [{\citenamefont {Xia}\ \emph {et~al.}(2015)\citenamefont {Xia}, \citenamefont {Zhao}, \citenamefont {Twamley},\ and\ \citenamefont {{EQuS Collaboration}}}]{xia_detection_2015}%
  \BibitemOpen
  \bibfield  {author} {\bibinfo {author} {\bibfnamefont {K.}~\bibnamefont {Xia}}, \bibinfo {author} {\bibfnamefont {N.}~\bibnamefont {Zhao}}, \bibinfo {author} {\bibfnamefont {J.}~\bibnamefont {Twamley}},\ and\ \bibinfo {author} {\bibnamefont {{EQuS Collaboration}}},\ }\bibfield  {title} {\bibinfo {title} {Detection of a weak magnetic field via cavity-enhanced {Faraday} rotation},\ }\href {https://doi.org/10.1103/PhysRevA.92.043409} {\bibfield  {journal} {\bibinfo  {journal} {Physical Review A}\ }\textbf {\bibinfo {volume} {92}},\ \bibinfo {pages} {043409} (\bibinfo {year} {2015})}\BibitemShut {NoStop}%
\bibitem [{\citenamefont {Kuiri}\ \emph {et~al.}(2022)\citenamefont {Kuiri}, \citenamefont {Coleman}, \citenamefont {Gao}, \citenamefont {Vishnuradhan}, \citenamefont {Watanabe}, \citenamefont {Taniguchi}, \citenamefont {Zhu}, \citenamefont {MacDonald},\ and\ \citenamefont {Folk}}]{corr_kuiri_spontaneous_2022}%
  \BibitemOpen
  \bibfield  {author} {\bibinfo {author} {\bibfnamefont {M.}~\bibnamefont {Kuiri}}, \bibinfo {author} {\bibfnamefont {C.}~\bibnamefont {Coleman}}, \bibinfo {author} {\bibfnamefont {Z.}~\bibnamefont {Gao}}, \bibinfo {author} {\bibfnamefont {A.}~\bibnamefont {Vishnuradhan}}, \bibinfo {author} {\bibfnamefont {K.}~\bibnamefont {Watanabe}}, \bibinfo {author} {\bibfnamefont {T.}~\bibnamefont {Taniguchi}}, \bibinfo {author} {\bibfnamefont {J.}~\bibnamefont {Zhu}}, \bibinfo {author} {\bibfnamefont {A.~H.}\ \bibnamefont {MacDonald}},\ and\ \bibinfo {author} {\bibfnamefont {J.}~\bibnamefont {Folk}},\ }\bibfield  {title} {\bibinfo {title} {Spontaneous time-reversal symmetry breaking in twisted double bilayer graphene},\ }\href {https://doi.org/10.1038/s41467-022-34192-x} {\bibfield  {journal} {\bibinfo  {journal} {Nature Communications}\ }\textbf {\bibinfo {volume} {13}},\ \bibinfo {pages} {6468} (\bibinfo {year} {2022})}\BibitemShut {NoStop}%
\bibitem [{\citenamefont {Lee}\ \emph {et~al.}(2019)\citenamefont {Lee}, \citenamefont {Khalaf}, \citenamefont {Liu}, \citenamefont {Liu}, \citenamefont {Hao}, \citenamefont {Kim},\ and\ \citenamefont {Vishwanath}}]{corr_lee_theory_2019}%
  \BibitemOpen
  \bibfield  {author} {\bibinfo {author} {\bibfnamefont {J.~Y.}\ \bibnamefont {Lee}}, \bibinfo {author} {\bibfnamefont {E.}~\bibnamefont {Khalaf}}, \bibinfo {author} {\bibfnamefont {S.}~\bibnamefont {Liu}}, \bibinfo {author} {\bibfnamefont {X.}~\bibnamefont {Liu}}, \bibinfo {author} {\bibfnamefont {Z.}~\bibnamefont {Hao}}, \bibinfo {author} {\bibfnamefont {P.}~\bibnamefont {Kim}},\ and\ \bibinfo {author} {\bibfnamefont {A.}~\bibnamefont {Vishwanath}},\ }\bibfield  {title} {\bibinfo {title} {Theory of correlated insulating behaviour and spin-triplet superconductivity in twisted double bilayer graphene},\ }\href {https://doi.org/10.1038/s41467-019-12981-1} {\bibfield  {journal} {\bibinfo  {journal} {Nature Communications}\ }\textbf {\bibinfo {volume} {10}},\ \bibinfo {pages} {5333} (\bibinfo {year} {2019})}\BibitemShut {NoStop}%
\bibitem [{\citenamefont {Liu}\ \emph {et~al.}(2020)\citenamefont {Liu}, \citenamefont {Hao}, \citenamefont {Khalaf}, \citenamefont {Lee}, \citenamefont {Ronen}, \citenamefont {Yoo}, \citenamefont {Haei~Najafabadi}, \citenamefont {Watanabe}, \citenamefont {Taniguchi}, \citenamefont {Vishwanath},\ and\ \citenamefont {Kim}}]{corr_liu_tunable_2020}%
  \BibitemOpen
  \bibfield  {author} {\bibinfo {author} {\bibfnamefont {X.}~\bibnamefont {Liu}}, \bibinfo {author} {\bibfnamefont {Z.}~\bibnamefont {Hao}}, \bibinfo {author} {\bibfnamefont {E.}~\bibnamefont {Khalaf}}, \bibinfo {author} {\bibfnamefont {J.~Y.}\ \bibnamefont {Lee}}, \bibinfo {author} {\bibfnamefont {Y.}~\bibnamefont {Ronen}}, \bibinfo {author} {\bibfnamefont {H.}~\bibnamefont {Yoo}}, \bibinfo {author} {\bibfnamefont {D.}~\bibnamefont {Haei~Najafabadi}}, \bibinfo {author} {\bibfnamefont {K.}~\bibnamefont {Watanabe}}, \bibinfo {author} {\bibfnamefont {T.}~\bibnamefont {Taniguchi}}, \bibinfo {author} {\bibfnamefont {A.}~\bibnamefont {Vishwanath}},\ and\ \bibinfo {author} {\bibfnamefont {P.}~\bibnamefont {Kim}},\ }\bibfield  {title} {\bibinfo {title} {Tunable spin-polarized correlated states in twisted double bilayer graphene},\ }\href {https://doi.org/10.1038/s41586-020-2458-7} {\bibfield  {journal} {\bibinfo  {journal} {Nature}\ }\textbf {\bibinfo {volume} {583}},\ \bibinfo {pages} {221} (\bibinfo {year}
  {2020})}\BibitemShut {NoStop}%
\bibitem [{\citenamefont {Ghosh}\ \emph {et~al.}(2020)\citenamefont {Ghosh}, \citenamefont {Smidman}, \citenamefont {Shang}, \citenamefont {Annett}, \citenamefont {Hillier}, \citenamefont {Quintanilla},\ and\ \citenamefont {Yuan}}]{ghosh2020recent}%
  \BibitemOpen
  \bibfield  {author} {\bibinfo {author} {\bibfnamefont {S.~K.}\ \bibnamefont {Ghosh}}, \bibinfo {author} {\bibfnamefont {M.}~\bibnamefont {Smidman}}, \bibinfo {author} {\bibfnamefont {T.}~\bibnamefont {Shang}}, \bibinfo {author} {\bibfnamefont {J.~F.}\ \bibnamefont {Annett}}, \bibinfo {author} {\bibfnamefont {A.~D.}\ \bibnamefont {Hillier}}, \bibinfo {author} {\bibfnamefont {J.}~\bibnamefont {Quintanilla}},\ and\ \bibinfo {author} {\bibfnamefont {H.}~\bibnamefont {Yuan}},\ }\bibfield  {title} {\bibinfo {title} {Recent progress on superconductors with time-reversal symmetry breaking},\ }\href@noop {} {\bibfield  {journal} {\bibinfo  {journal} {Journal of Physics: Condensed Matter}\ }\textbf {\bibinfo {volume} {33}},\ \bibinfo {pages} {033001} (\bibinfo {year} {2020})}\BibitemShut {NoStop}%
\bibitem [{\citenamefont {Read}\ and\ \citenamefont {Green}(2000)}]{Read.2000}%
  \BibitemOpen
  \bibfield  {author} {\bibinfo {author} {\bibfnamefont {N.}~\bibnamefont {Read}}\ and\ \bibinfo {author} {\bibfnamefont {D.}~\bibnamefont {Green}},\ }\bibfield  {title} {\bibinfo {title} {{Paired states of fermions in two dimensions with breaking of parity and time-reversal symmetries and the fractional quantum Hall effect}},\ }\href {https://doi.org/10.1103/physrevb.61.10267} {\bibfield  {journal} {\bibinfo  {journal} {Physical Review B}\ }\textbf {\bibinfo {volume} {61}},\ \bibinfo {pages} {10267} (\bibinfo {year} {2000})}\BibitemShut {NoStop}%
\bibitem [{\citenamefont {Kallin}\ and\ \citenamefont {Berlinsky}(2016)}]{kallin_chiral_2016}%
  \BibitemOpen
  \bibfield  {author} {\bibinfo {author} {\bibfnamefont {C.}~\bibnamefont {Kallin}}\ and\ \bibinfo {author} {\bibfnamefont {J.}~\bibnamefont {Berlinsky}},\ }\bibfield  {title} {\bibinfo {title} {Chiral superconductors},\ }\href {https://doi.org/10.1088/0034-4885/79/5/054502} {\bibfield  {journal} {\bibinfo  {journal} {Reports on Progress in Physics}\ }\textbf {\bibinfo {volume} {79}},\ \bibinfo {pages} {054502} (\bibinfo {year} {2016})}\BibitemShut {NoStop}%
\bibitem [{\citenamefont {Poniatowski}\ \emph {et~al.}(2021)\citenamefont {Poniatowski}, \citenamefont {Curtis}, \citenamefont {Yacoby},\ and\ \citenamefont {Narang}}]{Poniatowski.2022a}%
  \BibitemOpen
  \bibfield  {author} {\bibinfo {author} {\bibfnamefont {N.}~\bibnamefont {Poniatowski}}, \bibinfo {author} {\bibfnamefont {J.}~\bibnamefont {Curtis}}, \bibinfo {author} {\bibfnamefont {A.}~\bibnamefont {Yacoby}},\ and\ \bibinfo {author} {\bibfnamefont {P.}~\bibnamefont {Narang}},\ }\bibfield  {title} {\bibinfo {title} {Spectroscopic signatures of time-reversal symmetry breaking superconductivity},\ }\href {https://doi.org/10.1038/s42005-022-00819-0} {\bibfield  {journal} {\bibinfo  {journal} {Comm. Phys.}\ }\textbf {\bibinfo {volume} {5}},\ \bibinfo {pages} {44} (\bibinfo {year} {2021})}\BibitemShut {NoStop}%
\bibitem [{\citenamefont {Curtis}\ \emph {et~al.}(2022)\citenamefont {Curtis}, \citenamefont {Poniatowski}, \citenamefont {Yacoby},\ and\ \citenamefont {Narang}}]{Curtis.2022a}%
  \BibitemOpen
  \bibfield  {author} {\bibinfo {author} {\bibfnamefont {J.}~\bibnamefont {Curtis}}, \bibinfo {author} {\bibfnamefont {N.}~\bibnamefont {Poniatowski}}, \bibinfo {author} {\bibfnamefont {A.}~\bibnamefont {Yacoby}},\ and\ \bibinfo {author} {\bibfnamefont {P.}~\bibnamefont {Narang}},\ }\bibfield  {title} {\bibinfo {title} {Proximity-induced collective modes in an unconventional superconductor heterostructure},\ }\href {https://doi.org/10.1103/PhysRevB.106.064508} {\bibfield  {journal} {\bibinfo  {journal} {Phys. Rev. B}\ }\textbf {\bibinfo {volume} {106}},\ \bibinfo {pages} {064508} (\bibinfo {year} {2022})}\BibitemShut {NoStop}%
\bibitem [{\citenamefont {Basov}\ \emph {et~al.}(2011)\citenamefont {Basov}, \citenamefont {Averitt}, \citenamefont {Van Der~Marel}, \citenamefont {Dressel},\ and\ \citenamefont {Haule}}]{basov2011electrodynamics}%
  \BibitemOpen
  \bibfield  {author} {\bibinfo {author} {\bibfnamefont {D.~N.}\ \bibnamefont {Basov}}, \bibinfo {author} {\bibfnamefont {R.~D.}\ \bibnamefont {Averitt}}, \bibinfo {author} {\bibfnamefont {D.}~\bibnamefont {Van Der~Marel}}, \bibinfo {author} {\bibfnamefont {M.}~\bibnamefont {Dressel}},\ and\ \bibinfo {author} {\bibfnamefont {K.}~\bibnamefont {Haule}},\ }\bibfield  {title} {\bibinfo {title} {Electrodynamics of correlated electron materials},\ }\href@noop {} {\bibfield  {journal} {\bibinfo  {journal} {Reviews of Modern Physics}\ }\textbf {\bibinfo {volume} {83}},\ \bibinfo {pages} {471} (\bibinfo {year} {2011})}\BibitemShut {NoStop}%
\bibitem [{\citenamefont {Han}\ \emph {et~al.}(2025)\citenamefont {Han}, \citenamefont {Lu}, \citenamefont {Hadjri}, \citenamefont {Shi}, \citenamefont {Wu}, \citenamefont {Xu}, \citenamefont {Yao}, \citenamefont {Cotten}, \citenamefont {Sedeh}, \citenamefont {Weldeyesus} \emph {et~al.}}]{han2025signatures}%
  \BibitemOpen
  \bibfield  {author} {\bibinfo {author} {\bibfnamefont {T.}~\bibnamefont {Han}}, \bibinfo {author} {\bibfnamefont {Z.}~\bibnamefont {Lu}}, \bibinfo {author} {\bibfnamefont {Z.}~\bibnamefont {Hadjri}}, \bibinfo {author} {\bibfnamefont {L.}~\bibnamefont {Shi}}, \bibinfo {author} {\bibfnamefont {Z.}~\bibnamefont {Wu}}, \bibinfo {author} {\bibfnamefont {W.}~\bibnamefont {Xu}}, \bibinfo {author} {\bibfnamefont {Y.}~\bibnamefont {Yao}}, \bibinfo {author} {\bibfnamefont {A.~A.}\ \bibnamefont {Cotten}}, \bibinfo {author} {\bibfnamefont {O.~S.}\ \bibnamefont {Sedeh}}, \bibinfo {author} {\bibfnamefont {H.}~\bibnamefont {Weldeyesus}}, \emph {et~al.},\ }\bibfield  {title} {\bibinfo {title} {Signatures of chiral superconductivity in rhombohedral graphene},\ }\href@noop {} {\bibfield  {journal} {\bibinfo  {journal} {Nature}\ ,\ \bibinfo {pages} {1}} (\bibinfo {year} {2025})}\BibitemShut {NoStop}%
\bibitem [{\citenamefont {Parra-Martinez}\ \emph {et~al.}(2025)\citenamefont {Parra-Martinez}, \citenamefont {Jimeno-Pozo}, \citenamefont {Phong}, \citenamefont {Sainz-Cruz}, \citenamefont {Kaplan}, \citenamefont {Emanuel}, \citenamefont {Oreg}, \citenamefont {Pantale{\'o}n}, \citenamefont {Silva-Guill{\'e}n},\ and\ \citenamefont {Guinea}}]{parra2025band}%
  \BibitemOpen
  \bibfield  {author} {\bibinfo {author} {\bibfnamefont {G.}~\bibnamefont {Parra-Martinez}}, \bibinfo {author} {\bibfnamefont {A.}~\bibnamefont {Jimeno-Pozo}}, \bibinfo {author} {\bibfnamefont {V.~T.}\ \bibnamefont {Phong}}, \bibinfo {author} {\bibfnamefont {H.}~\bibnamefont {Sainz-Cruz}}, \bibinfo {author} {\bibfnamefont {D.}~\bibnamefont {Kaplan}}, \bibinfo {author} {\bibfnamefont {P.}~\bibnamefont {Emanuel}}, \bibinfo {author} {\bibfnamefont {Y.}~\bibnamefont {Oreg}}, \bibinfo {author} {\bibfnamefont {P.~A.}\ \bibnamefont {Pantale{\'o}n}}, \bibinfo {author} {\bibfnamefont {J.~{\'A}.}\ \bibnamefont {Silva-Guill{\'e}n}},\ and\ \bibinfo {author} {\bibfnamefont {F.}~\bibnamefont {Guinea}},\ }\bibfield  {title} {\bibinfo {title} {Band renormalization, quarter metals, and chiral superconductivity in rhombohedral tetralayer graphene},\ }\href@noop {} {\bibfield  {journal} {\bibinfo  {journal} {Physical Review Letters}\ }\textbf {\bibinfo {volume} {135}},\ \bibinfo {pages} {136503} (\bibinfo {year}
  {2025})}\BibitemShut {NoStop}%
\bibitem [{\citenamefont {Petrides}\ and\ \citenamefont {Zilberberg}(2022)}]{petrides2022semiclassical}%
  \BibitemOpen
  \bibfield  {author} {\bibinfo {author} {\bibfnamefont {I.}~\bibnamefont {Petrides}}\ and\ \bibinfo {author} {\bibfnamefont {O.}~\bibnamefont {Zilberberg}},\ }\bibfield  {title} {\bibinfo {title} {Semiclassical treatment of spinor topological effects in driven inhomogeneous insulators under external electromagnetic fields},\ }\href@noop {} {\bibfield  {journal} {\bibinfo  {journal} {Physical Review B}\ }\textbf {\bibinfo {volume} {106}},\ \bibinfo {pages} {165130} (\bibinfo {year} {2022})}\BibitemShut {NoStop}%
\bibitem [{\citenamefont {Zhao}\ \emph {et~al.}(2020)\citenamefont {Zhao}, \citenamefont {Guo}, \citenamefont {Garcia}, \citenamefont {Narang},\ and\ \citenamefont {Fan}}]{zhao2020axion}%
  \BibitemOpen
  \bibfield  {author} {\bibinfo {author} {\bibfnamefont {B.}~\bibnamefont {Zhao}}, \bibinfo {author} {\bibfnamefont {C.}~\bibnamefont {Guo}}, \bibinfo {author} {\bibfnamefont {C.~A.}\ \bibnamefont {Garcia}}, \bibinfo {author} {\bibfnamefont {P.}~\bibnamefont {Narang}},\ and\ \bibinfo {author} {\bibfnamefont {S.}~\bibnamefont {Fan}},\ }\bibfield  {title} {\bibinfo {title} {Axion-field-enabled nonreciprocal thermal radiation in weyl semimetals},\ }\href@noop {} {\bibfield  {journal} {\bibinfo  {journal} {Nano letters}\ }\textbf {\bibinfo {volume} {20}},\ \bibinfo {pages} {1923} (\bibinfo {year} {2020})}\BibitemShut {NoStop}%
\bibitem [{\citenamefont {Curtis}\ \emph {et~al.}(2023)\citenamefont {Curtis}, \citenamefont {Petrides},\ and\ \citenamefont {Narang}}]{Curtis.2023a}%
  \BibitemOpen
  \bibfield  {author} {\bibinfo {author} {\bibfnamefont {J.}~\bibnamefont {Curtis}}, \bibinfo {author} {\bibfnamefont {I.}~\bibnamefont {Petrides}},\ and\ \bibinfo {author} {\bibfnamefont {P.}~\bibnamefont {Narang}},\ }\bibfield  {title} {\bibinfo {title} {Finite-momentum instability of dynamical axion insulator},\ }\href {https://doi.org/xxxxxx} {\bibfield  {journal} {\bibinfo  {journal} {Phys. Rev. B}\ }\textbf {\bibinfo {volume} {xxx}},\ \bibinfo {pages} {xxxxx} (\bibinfo {year} {2023})}\BibitemShut {NoStop}%
\bibitem [{\citenamefont {Narang}\ \emph {et~al.}(2021)\citenamefont {Narang}, \citenamefont {Garcia},\ and\ \citenamefont {Felser}}]{Narang.2021}%
  \BibitemOpen
  \bibfield  {author} {\bibinfo {author} {\bibfnamefont {P.}~\bibnamefont {Narang}}, \bibinfo {author} {\bibfnamefont {C.}~\bibnamefont {Garcia}},\ and\ \bibinfo {author} {\bibfnamefont {C.}~\bibnamefont {Felser}},\ }\bibfield  {title} {\bibinfo {title} {The topology of electronic band structures},\ }\href {https://doi.org/10.1038/s41563-020-00820-4} {\bibfield  {journal} {\bibinfo  {journal} {Nature Mat.}\ }\textbf {\bibinfo {volume} {20}},\ \bibinfo {pages} {293} (\bibinfo {year} {2021})}\BibitemShut {NoStop}%
\bibitem [{\citenamefont {Nichol}\ \emph {et~al.}(2017)\citenamefont {Nichol}, \citenamefont {Orona}, \citenamefont {Harvey}, \citenamefont {Fallahi}, \citenamefont {Gardner}, \citenamefont {Manfra},\ and\ \citenamefont {Yacoby}}]{nichol2017high}%
  \BibitemOpen
  \bibfield  {author} {\bibinfo {author} {\bibfnamefont {J.~M.}\ \bibnamefont {Nichol}}, \bibinfo {author} {\bibfnamefont {L.~A.}\ \bibnamefont {Orona}}, \bibinfo {author} {\bibfnamefont {S.~P.}\ \bibnamefont {Harvey}}, \bibinfo {author} {\bibfnamefont {S.}~\bibnamefont {Fallahi}}, \bibinfo {author} {\bibfnamefont {G.~C.}\ \bibnamefont {Gardner}}, \bibinfo {author} {\bibfnamefont {M.~J.}\ \bibnamefont {Manfra}},\ and\ \bibinfo {author} {\bibfnamefont {A.}~\bibnamefont {Yacoby}},\ }\bibfield  {title} {\bibinfo {title} {High-fidelity entangling gate for double-quantum-dot spin qubits},\ }\href@noop {} {\bibfield  {journal} {\bibinfo  {journal} {npj Quantum Information}\ }\textbf {\bibinfo {volume} {3}},\ \bibinfo {pages} {3} (\bibinfo {year} {2017})}\BibitemShut {NoStop}%
\bibitem [{\citenamefont {Takahashi}\ \emph {et~al.}(2013)\citenamefont {Takahashi}, \citenamefont {Bartlett},\ and\ \citenamefont {Doherty}}]{takahashi2013tomography}%
  \BibitemOpen
  \bibfield  {author} {\bibinfo {author} {\bibfnamefont {M.}~\bibnamefont {Takahashi}}, \bibinfo {author} {\bibfnamefont {S.~D.}\ \bibnamefont {Bartlett}},\ and\ \bibinfo {author} {\bibfnamefont {A.~C.}\ \bibnamefont {Doherty}},\ }\bibfield  {title} {\bibinfo {title} {Tomography of a spin qubit in a double quantum dot},\ }\href@noop {} {\bibfield  {journal} {\bibinfo  {journal} {Physical Review A}\ }\textbf {\bibinfo {volume} {88}},\ \bibinfo {pages} {022120} (\bibinfo {year} {2013})}\BibitemShut {NoStop}%
\bibitem [{\citenamefont {Yu}\ \emph {et~al.}(2022)\citenamefont {Yu}, \citenamefont {Liu}, \citenamefont {Yang}, \citenamefont {Gong}, \citenamefont {Cao}, \citenamefont {Zhang}, \citenamefont {Liu}, \citenamefont {Heyl}, \citenamefont {Ozawa}, \citenamefont {Goldman} \emph {et~al.}}]{yu2022quantum}%
  \BibitemOpen
  \bibfield  {author} {\bibinfo {author} {\bibfnamefont {M.}~\bibnamefont {Yu}}, \bibinfo {author} {\bibfnamefont {Y.}~\bibnamefont {Liu}}, \bibinfo {author} {\bibfnamefont {P.}~\bibnamefont {Yang}}, \bibinfo {author} {\bibfnamefont {M.}~\bibnamefont {Gong}}, \bibinfo {author} {\bibfnamefont {Q.}~\bibnamefont {Cao}}, \bibinfo {author} {\bibfnamefont {S.}~\bibnamefont {Zhang}}, \bibinfo {author} {\bibfnamefont {H.}~\bibnamefont {Liu}}, \bibinfo {author} {\bibfnamefont {M.}~\bibnamefont {Heyl}}, \bibinfo {author} {\bibfnamefont {T.}~\bibnamefont {Ozawa}}, \bibinfo {author} {\bibfnamefont {N.}~\bibnamefont {Goldman}}, \emph {et~al.},\ }\bibfield  {title} {\bibinfo {title} {Quantum fisher information measurement and verification of the quantum cram{\'e}r--rao bound in a solid-state qubit},\ }\href@noop {} {\bibfield  {journal} {\bibinfo  {journal} {npj Quantum Information}\ }\textbf {\bibinfo {volume} {8}},\ \bibinfo {pages} {56} (\bibinfo {year} {2022})}\BibitemShut {NoStop}%
\bibitem [{\citenamefont {Hofheinz}\ \emph {et~al.}(2008)\citenamefont {Hofheinz}, \citenamefont {Weig}, \citenamefont {Ansmann}, \citenamefont {Bialczak}, \citenamefont {Lucero}, \citenamefont {Neeley}, \citenamefont {O’Connell}, \citenamefont {Wang}, \citenamefont {Martinis},\ and\ \citenamefont {Cleland}}]{hofheinz_generation_2008}%
  \BibitemOpen
  \bibfield  {author} {\bibinfo {author} {\bibfnamefont {M.}~\bibnamefont {Hofheinz}}, \bibinfo {author} {\bibfnamefont {E.~M.}\ \bibnamefont {Weig}}, \bibinfo {author} {\bibfnamefont {M.}~\bibnamefont {Ansmann}}, \bibinfo {author} {\bibfnamefont {R.~C.}\ \bibnamefont {Bialczak}}, \bibinfo {author} {\bibfnamefont {E.}~\bibnamefont {Lucero}}, \bibinfo {author} {\bibfnamefont {M.}~\bibnamefont {Neeley}}, \bibinfo {author} {\bibfnamefont {A.~D.}\ \bibnamefont {O’Connell}}, \bibinfo {author} {\bibfnamefont {H.}~\bibnamefont {Wang}}, \bibinfo {author} {\bibfnamefont {J.~M.}\ \bibnamefont {Martinis}},\ and\ \bibinfo {author} {\bibfnamefont {A.~N.}\ \bibnamefont {Cleland}},\ }\bibfield  {title} {\bibinfo {title} {Generation of {Fock} states in a superconducting quantum circuit},\ }\href {https://doi.org/10.1038/nature07136} {\bibfield  {journal} {\bibinfo  {journal} {Nature}\ }\textbf {\bibinfo {volume} {454}},\ \bibinfo {pages} {310} (\bibinfo {year} {2008})}\BibitemShut {NoStop}%
\bibitem [{\citenamefont {Devoret}\ and\ \citenamefont {Martinis}(2005)}]{devoret2005implementing}%
  \BibitemOpen
  \bibfield  {author} {\bibinfo {author} {\bibfnamefont {M.~H.}\ \bibnamefont {Devoret}}\ and\ \bibinfo {author} {\bibfnamefont {J.~M.}\ \bibnamefont {Martinis}},\ }\bibfield  {title} {\bibinfo {title} {Implementing qubits with superconducting integrated circuits},\ }\href@noop {} {\bibfield  {journal} {\bibinfo  {journal} {Experimental aspects of quantum computing}\ ,\ \bibinfo {pages} {163}} (\bibinfo {year} {2005})}\BibitemShut {NoStop}%
\bibitem [{\citenamefont {Vallés-Sanclemente}\ \emph {et~al.}(2023)\citenamefont {Vallés-Sanclemente}, \citenamefont {van~der Meer}, \citenamefont {Finkel}, \citenamefont {Muthusubramanian}, \citenamefont {Beekman}, \citenamefont {Ali}, \citenamefont {Marques}, \citenamefont {Zachariadis}, \citenamefont {Veen}, \citenamefont {Stavenga}, \citenamefont {Haider},\ and\ \citenamefont {DiCarlo}}]{valles2023trimming}%
  \BibitemOpen
  \bibfield  {author} {\bibinfo {author} {\bibfnamefont {S.}~\bibnamefont {Vallés-Sanclemente}}, \bibinfo {author} {\bibfnamefont {S.~L.~M.}\ \bibnamefont {van~der Meer}}, \bibinfo {author} {\bibfnamefont {M.}~\bibnamefont {Finkel}}, \bibinfo {author} {\bibfnamefont {N.}~\bibnamefont {Muthusubramanian}}, \bibinfo {author} {\bibfnamefont {M.}~\bibnamefont {Beekman}}, \bibinfo {author} {\bibfnamefont {H.}~\bibnamefont {Ali}}, \bibinfo {author} {\bibfnamefont {J.~F.}\ \bibnamefont {Marques}}, \bibinfo {author} {\bibfnamefont {C.}~\bibnamefont {Zachariadis}}, \bibinfo {author} {\bibfnamefont {H.~M.}\ \bibnamefont {Veen}}, \bibinfo {author} {\bibfnamefont {T.}~\bibnamefont {Stavenga}}, \bibinfo {author} {\bibfnamefont {N.}~\bibnamefont {Haider}},\ and\ \bibinfo {author} {\bibfnamefont {L.}~\bibnamefont {DiCarlo}},\ }\href {http://arxiv.org/abs/2302.10705} {\bibinfo {title} {Post-fabrication frequency trimming of coplanar-waveguide resonators in circuit {QED} quantum processors}} (\bibinfo {year} {2023}),\
  \bibinfo {note} {arXiv:2302.10705 [cond-mat, physics:quant-ph]}\BibitemShut {NoStop}%
\bibitem [{\citenamefont {Soltani}\ \emph {et~al.}(2017)\citenamefont {Soltani}, \citenamefont {Zhang}, \citenamefont {Ryan}, \citenamefont {Ribeill}, \citenamefont {Wang},\ and\ \citenamefont {Loncar}}]{soltani2017efficient}%
  \BibitemOpen
  \bibfield  {author} {\bibinfo {author} {\bibfnamefont {M.}~\bibnamefont {Soltani}}, \bibinfo {author} {\bibfnamefont {M.}~\bibnamefont {Zhang}}, \bibinfo {author} {\bibfnamefont {C.}~\bibnamefont {Ryan}}, \bibinfo {author} {\bibfnamefont {G.~J.}\ \bibnamefont {Ribeill}}, \bibinfo {author} {\bibfnamefont {C.}~\bibnamefont {Wang}},\ and\ \bibinfo {author} {\bibfnamefont {M.}~\bibnamefont {Loncar}},\ }\bibfield  {title} {\bibinfo {title} {Efficient quantum microwave-to-optical conversion using electro-optic nanophotonic coupled resonators},\ }\href@noop {} {\bibfield  {journal} {\bibinfo  {journal} {Physical Review A}\ }\textbf {\bibinfo {volume} {96}},\ \bibinfo {pages} {043808} (\bibinfo {year} {2017})}\BibitemShut {NoStop}%
\bibitem [{\citenamefont {Fan}\ \emph {et~al.}(2018)\citenamefont {Fan}, \citenamefont {Zou}, \citenamefont {Cheng}, \citenamefont {Guo}, \citenamefont {Han}, \citenamefont {Gong}, \citenamefont {Wang},\ and\ \citenamefont {Tang}}]{fan2018superconducting}%
  \BibitemOpen
  \bibfield  {author} {\bibinfo {author} {\bibfnamefont {L.}~\bibnamefont {Fan}}, \bibinfo {author} {\bibfnamefont {C.-L.}\ \bibnamefont {Zou}}, \bibinfo {author} {\bibfnamefont {R.}~\bibnamefont {Cheng}}, \bibinfo {author} {\bibfnamefont {X.}~\bibnamefont {Guo}}, \bibinfo {author} {\bibfnamefont {X.}~\bibnamefont {Han}}, \bibinfo {author} {\bibfnamefont {Z.}~\bibnamefont {Gong}}, \bibinfo {author} {\bibfnamefont {S.}~\bibnamefont {Wang}},\ and\ \bibinfo {author} {\bibfnamefont {H.~X.}\ \bibnamefont {Tang}},\ }\bibfield  {title} {\bibinfo {title} {Superconducting cavity electro-optics: a platform for coherent photon conversion between superconducting and photonic circuits},\ }\href@noop {} {\bibfield  {journal} {\bibinfo  {journal} {Science advances}\ }\textbf {\bibinfo {volume} {4}},\ \bibinfo {pages} {eaar4994} (\bibinfo {year} {2018})}\BibitemShut {NoStop}%
\bibitem [{\citenamefont {Petrides}\ \emph {et~al.}(2025)\citenamefont {Petrides}, \citenamefont {Curtis}, \citenamefont {Wesson}, \citenamefont {Yacoby},\ and\ \citenamefont {Narang}}]{petrides2025probing}%
  \BibitemOpen
  \bibfield  {author} {\bibinfo {author} {\bibfnamefont {I.}~\bibnamefont {Petrides}}, \bibinfo {author} {\bibfnamefont {J.~B.}\ \bibnamefont {Curtis}}, \bibinfo {author} {\bibfnamefont {M.}~\bibnamefont {Wesson}}, \bibinfo {author} {\bibfnamefont {A.}~\bibnamefont {Yacoby}},\ and\ \bibinfo {author} {\bibfnamefont {P.}~\bibnamefont {Narang}},\ }\bibfield  {title} {\bibinfo {title} {Probing electromagnetic nonreciprocity with quantum geometry of photonic states},\ }\href@noop {} {\bibfield  {journal} {\bibinfo  {journal} {Physical Review Research}\ }\textbf {\bibinfo {volume} {7}},\ \bibinfo {pages} {023006} (\bibinfo {year} {2025})}\BibitemShut {NoStop}%
\bibitem [{\citenamefont {Wang}\ \emph {et~al.}(2013)\citenamefont {Wang}, \citenamefont {Meric}, \citenamefont {Huang}, \citenamefont {Gao}, \citenamefont {Gao}, \citenamefont {Tran}, \citenamefont {Taniguchi}, \citenamefont {Watanabe}, \citenamefont {Campos}, \citenamefont {Muller} \emph {et~al.}}]{wang2013one}%
  \BibitemOpen
  \bibfield  {author} {\bibinfo {author} {\bibfnamefont {L.}~\bibnamefont {Wang}}, \bibinfo {author} {\bibfnamefont {I.}~\bibnamefont {Meric}}, \bibinfo {author} {\bibfnamefont {P.}~\bibnamefont {Huang}}, \bibinfo {author} {\bibfnamefont {Q.}~\bibnamefont {Gao}}, \bibinfo {author} {\bibfnamefont {Y.}~\bibnamefont {Gao}}, \bibinfo {author} {\bibfnamefont {H.}~\bibnamefont {Tran}}, \bibinfo {author} {\bibfnamefont {T.}~\bibnamefont {Taniguchi}}, \bibinfo {author} {\bibfnamefont {K.}~\bibnamefont {Watanabe}}, \bibinfo {author} {\bibfnamefont {L.}~\bibnamefont {Campos}}, \bibinfo {author} {\bibfnamefont {D.}~\bibnamefont {Muller}}, \emph {et~al.},\ }\bibfield  {title} {\bibinfo {title} {One-dimensional electrical contact to a two-dimensional material},\ }\href@noop {} {\bibfield  {journal} {\bibinfo  {journal} {Science}\ }\textbf {\bibinfo {volume} {342}},\ \bibinfo {pages} {614} (\bibinfo {year} {2013})}\BibitemShut {NoStop}%
\bibitem [{\citenamefont {Cao}\ \emph {et~al.}(2018)\citenamefont {Cao}, \citenamefont {Fatemi}, \citenamefont {Fang}, \citenamefont {Watanabe}, \citenamefont {Taniguchi}, \citenamefont {Kaxiras},\ and\ \citenamefont {Jarillo-Herrero}}]{cao_unconventional_2018}%
  \BibitemOpen
  \bibfield  {author} {\bibinfo {author} {\bibfnamefont {Y.}~\bibnamefont {Cao}}, \bibinfo {author} {\bibfnamefont {V.}~\bibnamefont {Fatemi}}, \bibinfo {author} {\bibfnamefont {S.}~\bibnamefont {Fang}}, \bibinfo {author} {\bibfnamefont {K.}~\bibnamefont {Watanabe}}, \bibinfo {author} {\bibfnamefont {T.}~\bibnamefont {Taniguchi}}, \bibinfo {author} {\bibfnamefont {E.}~\bibnamefont {Kaxiras}},\ and\ \bibinfo {author} {\bibfnamefont {P.}~\bibnamefont {Jarillo-Herrero}},\ }\bibfield  {title} {\bibinfo {title} {Unconventional superconductivity in magic-angle graphene superlattices},\ }\href {https://doi.org/10.1038/nature26160} {\bibfield  {journal} {\bibinfo  {journal} {Nature}\ }\textbf {\bibinfo {volume} {556}},\ \bibinfo {pages} {43} (\bibinfo {year} {2018})}\BibitemShut {NoStop}%
\bibitem [{\citenamefont {Sajadi}\ \emph {et~al.}(2018)\citenamefont {Sajadi}, \citenamefont {Palomaki}, \citenamefont {Fei}, \citenamefont {Zhao}, \citenamefont {Bement}, \citenamefont {Olsen}, \citenamefont {Luescher}, \citenamefont {Xu}, \citenamefont {Folk},\ and\ \citenamefont {Cobden}}]{sajadi_gate-induced_2018}%
  \BibitemOpen
  \bibfield  {author} {\bibinfo {author} {\bibfnamefont {E.}~\bibnamefont {Sajadi}}, \bibinfo {author} {\bibfnamefont {T.}~\bibnamefont {Palomaki}}, \bibinfo {author} {\bibfnamefont {Z.}~\bibnamefont {Fei}}, \bibinfo {author} {\bibfnamefont {W.}~\bibnamefont {Zhao}}, \bibinfo {author} {\bibfnamefont {P.}~\bibnamefont {Bement}}, \bibinfo {author} {\bibfnamefont {C.}~\bibnamefont {Olsen}}, \bibinfo {author} {\bibfnamefont {S.}~\bibnamefont {Luescher}}, \bibinfo {author} {\bibfnamefont {X.}~\bibnamefont {Xu}}, \bibinfo {author} {\bibfnamefont {J.~A.}\ \bibnamefont {Folk}},\ and\ \bibinfo {author} {\bibfnamefont {D.~H.}\ \bibnamefont {Cobden}},\ }\bibfield  {title} {\bibinfo {title} {Gate-induced superconductivity in a monolayer topological insulator},\ }\href {https://doi.org/10.1126/science.aar4426} {\bibfield  {journal} {\bibinfo  {journal} {Science}\ }\textbf {\bibinfo {volume} {362}},\ \bibinfo {pages} {922} (\bibinfo {year} {2018})}\BibitemShut {NoStop}%
\bibitem [{\citenamefont {Vallabhapurapu}\ \emph {et~al.}(2023)\citenamefont {Vallabhapurapu}, \citenamefont {Hansen}, \citenamefont {Adambukulam}, \citenamefont {St{\"o}hr}, \citenamefont {Denisenko}, \citenamefont {Yang},\ and\ \citenamefont {Laucht}}]{vallabhapurapu2023high}%
  \BibitemOpen
  \bibfield  {author} {\bibinfo {author} {\bibfnamefont {H.~H.}\ \bibnamefont {Vallabhapurapu}}, \bibinfo {author} {\bibfnamefont {I.}~\bibnamefont {Hansen}}, \bibinfo {author} {\bibfnamefont {C.}~\bibnamefont {Adambukulam}}, \bibinfo {author} {\bibfnamefont {R.}~\bibnamefont {St{\"o}hr}}, \bibinfo {author} {\bibfnamefont {A.}~\bibnamefont {Denisenko}}, \bibinfo {author} {\bibfnamefont {C.~H.}\ \bibnamefont {Yang}},\ and\ \bibinfo {author} {\bibfnamefont {A.}~\bibnamefont {Laucht}},\ }\bibfield  {title} {\bibinfo {title} {High-fidelity control of a nitrogen-vacancy-center spin qubit at room temperature using the sinusoidally modulated, always rotating, and tailored protocol},\ }\href@noop {} {\bibfield  {journal} {\bibinfo  {journal} {Physical Review A}\ }\textbf {\bibinfo {volume} {108}},\ \bibinfo {pages} {022606} (\bibinfo {year} {2023})}\BibitemShut {NoStop}%
\bibitem [{\citenamefont {Childress}\ and\ \citenamefont {Hanson}(2013)}]{childress2013diamond}%
  \BibitemOpen
  \bibfield  {author} {\bibinfo {author} {\bibfnamefont {L.}~\bibnamefont {Childress}}\ and\ \bibinfo {author} {\bibfnamefont {R.}~\bibnamefont {Hanson}},\ }\bibfield  {title} {\bibinfo {title} {Diamond nv centers for quantum computing and quantum networks},\ }\href@noop {} {\bibfield  {journal} {\bibinfo  {journal} {MRS bulletin}\ }\textbf {\bibinfo {volume} {38}},\ \bibinfo {pages} {134} (\bibinfo {year} {2013})}\BibitemShut {NoStop}%
\bibitem [{\citenamefont {Wang}\ \emph {et~al.}(2020)\citenamefont {Wang}, \citenamefont {Xiao}, \citenamefont {Liu}, \citenamefont {Lee-Wong}, \citenamefont {McLaughlin}, \citenamefont {Wang}, \citenamefont {Wu}, \citenamefont {Wang}, \citenamefont {Fullerton},\ and\ \citenamefont {Du}}]{wang2020electrical}%
  \BibitemOpen
  \bibfield  {author} {\bibinfo {author} {\bibfnamefont {X.}~\bibnamefont {Wang}}, \bibinfo {author} {\bibfnamefont {Y.}~\bibnamefont {Xiao}}, \bibinfo {author} {\bibfnamefont {C.}~\bibnamefont {Liu}}, \bibinfo {author} {\bibfnamefont {E.}~\bibnamefont {Lee-Wong}}, \bibinfo {author} {\bibfnamefont {N.~J.}\ \bibnamefont {McLaughlin}}, \bibinfo {author} {\bibfnamefont {H.}~\bibnamefont {Wang}}, \bibinfo {author} {\bibfnamefont {M.}~\bibnamefont {Wu}}, \bibinfo {author} {\bibfnamefont {H.}~\bibnamefont {Wang}}, \bibinfo {author} {\bibfnamefont {E.~E.}\ \bibnamefont {Fullerton}},\ and\ \bibinfo {author} {\bibfnamefont {C.~R.}\ \bibnamefont {Du}},\ }\bibfield  {title} {\bibinfo {title} {Electrical control of coherent spin rotation of a single-spin qubit},\ }\href@noop {} {\bibfield  {journal} {\bibinfo  {journal} {npj Quantum Information}\ }\textbf {\bibinfo {volume} {6}},\ \bibinfo {pages} {78} (\bibinfo {year} {2020})}\BibitemShut {NoStop}%
\bibitem [{\citenamefont {Oliver}\ \emph {et~al.}(2005)\citenamefont {Oliver}, \citenamefont {Yu}, \citenamefont {Lee}, \citenamefont {Berggren}, \citenamefont {Levitov},\ and\ \citenamefont {Orlando}}]{oliver2005mach}%
  \BibitemOpen
  \bibfield  {author} {\bibinfo {author} {\bibfnamefont {W.~D.}\ \bibnamefont {Oliver}}, \bibinfo {author} {\bibfnamefont {Y.}~\bibnamefont {Yu}}, \bibinfo {author} {\bibfnamefont {J.~C.}\ \bibnamefont {Lee}}, \bibinfo {author} {\bibfnamefont {K.~K.}\ \bibnamefont {Berggren}}, \bibinfo {author} {\bibfnamefont {L.~S.}\ \bibnamefont {Levitov}},\ and\ \bibinfo {author} {\bibfnamefont {T.~P.}\ \bibnamefont {Orlando}},\ }\bibfield  {title} {\bibinfo {title} {Mach-zehnder interferometry in a strongly driven superconducting qubit},\ }\href@noop {} {\bibfield  {journal} {\bibinfo  {journal} {Science}\ }\textbf {\bibinfo {volume} {310}},\ \bibinfo {pages} {1653} (\bibinfo {year} {2005})}\BibitemShut {NoStop}%
\bibitem [{\citenamefont {Khabiboulline}\ \emph {et~al.}(2019)\citenamefont {Khabiboulline}, \citenamefont {Borregaard}, \citenamefont {De~Greve},\ and\ \citenamefont {Lukin}}]{khabiboulline2019optical}%
  \BibitemOpen
  \bibfield  {author} {\bibinfo {author} {\bibfnamefont {E.~T.}\ \bibnamefont {Khabiboulline}}, \bibinfo {author} {\bibfnamefont {J.}~\bibnamefont {Borregaard}}, \bibinfo {author} {\bibfnamefont {K.}~\bibnamefont {De~Greve}},\ and\ \bibinfo {author} {\bibfnamefont {M.~D.}\ \bibnamefont {Lukin}},\ }\bibfield  {title} {\bibinfo {title} {Optical interferometry with quantum networks},\ }\href@noop {} {\bibfield  {journal} {\bibinfo  {journal} {Physical review letters}\ }\textbf {\bibinfo {volume} {123}},\ \bibinfo {pages} {070504} (\bibinfo {year} {2019})}\BibitemShut {NoStop}%
\bibitem [{\citenamefont {Cari{\~n}ena}\ \emph {et~al.}(2015)\citenamefont {Cari{\~n}ena}, \citenamefont {Ibort}, \citenamefont {Marmo}, \citenamefont {Morandi} \emph {et~al.}}]{carinena2015geometry}%
  \BibitemOpen
  \bibfield  {author} {\bibinfo {author} {\bibfnamefont {J.~F.}\ \bibnamefont {Cari{\~n}ena}}, \bibinfo {author} {\bibfnamefont {A.}~\bibnamefont {Ibort}}, \bibinfo {author} {\bibfnamefont {G.}~\bibnamefont {Marmo}}, \bibinfo {author} {\bibfnamefont {G.}~\bibnamefont {Morandi}}, \emph {et~al.},\ }\href@noop {} {\emph {\bibinfo {title} {Geometry from dynamics, classical and quantum}}}\ (\bibinfo  {publisher} {Springer},\ \bibinfo {year} {2015})\BibitemShut {NoStop}%
\bibitem [{\citenamefont {Ashtekar}\ and\ \citenamefont {Schilling}(1999)}]{ashtekar1999geometrical}%
  \BibitemOpen
  \bibfield  {author} {\bibinfo {author} {\bibfnamefont {A.}~\bibnamefont {Ashtekar}}\ and\ \bibinfo {author} {\bibfnamefont {T.~A.}\ \bibnamefont {Schilling}},\ }\bibfield  {title} {\bibinfo {title} {Geometrical formulation of quantum mechanics},\ }in\ \href@noop {} {\emph {\bibinfo {booktitle} {On Einstein’s Path: Essays in Honor of Engelbert Schucking}}}\ (\bibinfo  {publisher} {Springer},\ \bibinfo {year} {1999})\ pp.\ \bibinfo {pages} {23--65}\BibitemShut {NoStop}%
\bibitem [{\citenamefont {Phan}\ \emph {et~al.}(2022)\citenamefont {Phan}, \citenamefont {Senior}, \citenamefont {Ghazaryan}, \citenamefont {Hatefipour}, \citenamefont {Strickland}, \citenamefont {Shabani}, \citenamefont {Serbyn},\ and\ \citenamefont {Higginbotham}}]{phan2022detecting}%
  \BibitemOpen
  \bibfield  {author} {\bibinfo {author} {\bibfnamefont {D.}~\bibnamefont {Phan}}, \bibinfo {author} {\bibfnamefont {J.}~\bibnamefont {Senior}}, \bibinfo {author} {\bibfnamefont {A.}~\bibnamefont {Ghazaryan}}, \bibinfo {author} {\bibfnamefont {M.}~\bibnamefont {Hatefipour}}, \bibinfo {author} {\bibfnamefont {W.}~\bibnamefont {Strickland}}, \bibinfo {author} {\bibfnamefont {J.}~\bibnamefont {Shabani}}, \bibinfo {author} {\bibfnamefont {M.}~\bibnamefont {Serbyn}},\ and\ \bibinfo {author} {\bibfnamefont {A.~P.}\ \bibnamefont {Higginbotham}},\ }\bibfield  {title} {\bibinfo {title} {Detecting induced p$\pm$ip pairing at the al-inas interface with a quantum microwave circuit},\ }\href@noop {} {\bibfield  {journal} {\bibinfo  {journal} {Physical Review Letters}\ }\textbf {\bibinfo {volume} {128}},\ \bibinfo {pages} {107701} (\bibinfo {year} {2022})}\BibitemShut {NoStop}%
\bibitem [{\citenamefont {B{\o}ttcher}\ \emph {et~al.}(2024)\citenamefont {B{\o}ttcher}, \citenamefont {Poniatowski}, \citenamefont {Grankin}, \citenamefont {Wesson}, \citenamefont {Yan}, \citenamefont {Vool}, \citenamefont {Galitski},\ and\ \citenamefont {Yacoby}}]{bottcher2024circuit}%
  \BibitemOpen
  \bibfield  {author} {\bibinfo {author} {\bibfnamefont {C.}~\bibnamefont {B{\o}ttcher}}, \bibinfo {author} {\bibfnamefont {N.}~\bibnamefont {Poniatowski}}, \bibinfo {author} {\bibfnamefont {A.}~\bibnamefont {Grankin}}, \bibinfo {author} {\bibfnamefont {M.}~\bibnamefont {Wesson}}, \bibinfo {author} {\bibfnamefont {Z.}~\bibnamefont {Yan}}, \bibinfo {author} {\bibfnamefont {U.}~\bibnamefont {Vool}}, \bibinfo {author} {\bibfnamefont {V.}~\bibnamefont {Galitski}},\ and\ \bibinfo {author} {\bibfnamefont {A.}~\bibnamefont {Yacoby}},\ }\bibfield  {title} {\bibinfo {title} {Circuit quantum electrodynamics detection of induced two-fold anisotropic pairing in a hybrid superconductor--ferromagnet bilayer},\ }\href@noop {} {\bibfield  {journal} {\bibinfo  {journal} {Nature Physics}\ }\textbf {\bibinfo {volume} {20}},\ \bibinfo {pages} {1609} (\bibinfo {year} {2024})}\BibitemShut {NoStop}%
\bibitem [{\citenamefont {Tanaka}\ \emph {et~al.}(2024)\citenamefont {Tanaka}, \citenamefont {Wang}, \citenamefont {Dinh}, \citenamefont {Rodan-Legrain}, \citenamefont {Zaman}, \citenamefont {Hays}, \citenamefont {Kannan}, \citenamefont {Almanakly}, \citenamefont {Kim}, \citenamefont {Niedzielski} \emph {et~al.}}]{tanaka2024kinetic}%
  \BibitemOpen
  \bibfield  {author} {\bibinfo {author} {\bibfnamefont {M.}~\bibnamefont {Tanaka}}, \bibinfo {author} {\bibfnamefont {J.~{\^I}.}\ \bibnamefont {Wang}}, \bibinfo {author} {\bibfnamefont {T.~H.}\ \bibnamefont {Dinh}}, \bibinfo {author} {\bibfnamefont {D.}~\bibnamefont {Rodan-Legrain}}, \bibinfo {author} {\bibfnamefont {S.}~\bibnamefont {Zaman}}, \bibinfo {author} {\bibfnamefont {M.}~\bibnamefont {Hays}}, \bibinfo {author} {\bibfnamefont {B.}~\bibnamefont {Kannan}}, \bibinfo {author} {\bibfnamefont {A.}~\bibnamefont {Almanakly}}, \bibinfo {author} {\bibfnamefont {D.~K.}\ \bibnamefont {Kim}}, \bibinfo {author} {\bibfnamefont {B.~M.}\ \bibnamefont {Niedzielski}}, \emph {et~al.},\ }\bibfield  {title} {\bibinfo {title} {Kinetic inductance, quantum geometry, and superconductivity in magic-angle twisted bilayer graphene},\ }\href@noop {} {\bibfield  {journal} {\bibinfo  {journal} {arXiv preprint arXiv:2406.13740}\ } (\bibinfo {year} {2024})}\BibitemShut {NoStop}%
\bibitem [{\citenamefont {Banerjee}\ \emph {et~al.}(2024)\citenamefont {Banerjee}, \citenamefont {Hao}, \citenamefont {Kreidel}, \citenamefont {Ledwith}, \citenamefont {Phinney}, \citenamefont {Park}, \citenamefont {Zimmerman}, \citenamefont {Watanabe}, \citenamefont {Taniguchi}, \citenamefont {Westervelt} \emph {et~al.}}]{banerjee2024superfluid}%
  \BibitemOpen
  \bibfield  {author} {\bibinfo {author} {\bibfnamefont {A.}~\bibnamefont {Banerjee}}, \bibinfo {author} {\bibfnamefont {Z.}~\bibnamefont {Hao}}, \bibinfo {author} {\bibfnamefont {M.}~\bibnamefont {Kreidel}}, \bibinfo {author} {\bibfnamefont {P.}~\bibnamefont {Ledwith}}, \bibinfo {author} {\bibfnamefont {I.}~\bibnamefont {Phinney}}, \bibinfo {author} {\bibfnamefont {J.~M.}\ \bibnamefont {Park}}, \bibinfo {author} {\bibfnamefont {A.~M.}\ \bibnamefont {Zimmerman}}, \bibinfo {author} {\bibfnamefont {K.}~\bibnamefont {Watanabe}}, \bibinfo {author} {\bibfnamefont {T.}~\bibnamefont {Taniguchi}}, \bibinfo {author} {\bibfnamefont {R.~M.}\ \bibnamefont {Westervelt}}, \emph {et~al.},\ }\bibfield  {title} {\bibinfo {title} {Superfluid stiffness of twisted multilayer graphene superconductors},\ }\href@noop {} {\bibfield  {journal} {\bibinfo  {journal} {arXiv preprint arXiv:2406.13742}\ } (\bibinfo {year} {2024})}\BibitemShut {NoStop}%
\bibitem [{\citenamefont {Kreidel}\ \emph {et~al.}(2024)\citenamefont {Kreidel}, \citenamefont {Chu}, \citenamefont {Balgley}, \citenamefont {Antony}, \citenamefont {Verma}, \citenamefont {Ingham}, \citenamefont {Ranzani}, \citenamefont {Queiroz}, \citenamefont {Westervelt}, \citenamefont {Hone} \emph {et~al.}}]{kreidel2024measuring}%
  \BibitemOpen
  \bibfield  {author} {\bibinfo {author} {\bibfnamefont {M.}~\bibnamefont {Kreidel}}, \bibinfo {author} {\bibfnamefont {X.}~\bibnamefont {Chu}}, \bibinfo {author} {\bibfnamefont {J.}~\bibnamefont {Balgley}}, \bibinfo {author} {\bibfnamefont {A.}~\bibnamefont {Antony}}, \bibinfo {author} {\bibfnamefont {N.}~\bibnamefont {Verma}}, \bibinfo {author} {\bibfnamefont {J.}~\bibnamefont {Ingham}}, \bibinfo {author} {\bibfnamefont {L.}~\bibnamefont {Ranzani}}, \bibinfo {author} {\bibfnamefont {R.}~\bibnamefont {Queiroz}}, \bibinfo {author} {\bibfnamefont {R.~M.}\ \bibnamefont {Westervelt}}, \bibinfo {author} {\bibfnamefont {J.}~\bibnamefont {Hone}}, \emph {et~al.},\ }\bibfield  {title} {\bibinfo {title} {Measuring kinetic inductance and superfluid stiffness of two-dimensional superconductors using high-quality transmission-line resonators},\ }\href@noop {} {\bibfield  {journal} {\bibinfo  {journal} {Physical Review Research}\ }\textbf {\bibinfo {volume} {6}},\ \bibinfo {pages} {043245} (\bibinfo {year}
  {2024})}\BibitemShut {NoStop}%
\bibitem [{\citenamefont {Arora}\ and\ \citenamefont {Narang}(2025)}]{arora2025chiral}%
  \BibitemOpen
  \bibfield  {author} {\bibinfo {author} {\bibfnamefont {A.}~\bibnamefont {Arora}}\ and\ \bibinfo {author} {\bibfnamefont {P.}~\bibnamefont {Narang}},\ }\bibfield  {title} {\bibinfo {title} {Chiral cavity control of superconducting diode-like nonlinearities},\ }\href@noop {} {\bibfield  {journal} {\bibinfo  {journal} {arXiv preprint arXiv:2501.17924}\ } (\bibinfo {year} {2025})}\BibitemShut {NoStop}%
\bibitem [{\citenamefont {Arora}\ \emph {et~al.}(2025)\citenamefont {Arora}, \citenamefont {Curtis},\ and\ \citenamefont {Narang}}]{arora2025quantum}%
  \BibitemOpen
  \bibfield  {author} {\bibinfo {author} {\bibfnamefont {A.}~\bibnamefont {Arora}}, \bibinfo {author} {\bibfnamefont {J.~B.}\ \bibnamefont {Curtis}},\ and\ \bibinfo {author} {\bibfnamefont {P.}~\bibnamefont {Narang}},\ }\bibfield  {title} {\bibinfo {title} {Quantum geometry induced microwave enhancement of superconducting order in flat bands},\ }\href@noop {} {\bibfield  {journal} {\bibinfo  {journal} {Communications Physics}\ }\textbf {\bibinfo {volume} {8}},\ \bibinfo {pages} {327} (\bibinfo {year} {2025})}\BibitemShut {NoStop}%
\bibitem [{\citenamefont {Dirnegger}\ \emph {et~al.}(2025)\citenamefont {Dirnegger}, \citenamefont {Wesson}, \citenamefont {Arora}, \citenamefont {Petrides}, \citenamefont {Curtis}, \citenamefont {Been}, \citenamefont {Yacoby},\ and\ \citenamefont {Narang}}]{dirnegger2025nonlinear}%
  \BibitemOpen
  \bibfield  {author} {\bibinfo {author} {\bibfnamefont {N.}~\bibnamefont {Dirnegger}}, \bibinfo {author} {\bibfnamefont {M.}~\bibnamefont {Wesson}}, \bibinfo {author} {\bibfnamefont {A.}~\bibnamefont {Arora}}, \bibinfo {author} {\bibfnamefont {I.}~\bibnamefont {Petrides}}, \bibinfo {author} {\bibfnamefont {J.~B.}\ \bibnamefont {Curtis}}, \bibinfo {author} {\bibfnamefont {E.~M.}\ \bibnamefont {Been}}, \bibinfo {author} {\bibfnamefont {A.}~\bibnamefont {Yacoby}},\ and\ \bibinfo {author} {\bibfnamefont {P.}~\bibnamefont {Narang}},\ }\bibfield  {title} {\bibinfo {title} {Nonlinear superconducting ring resonator for sensitive measurement of time reversal symmetry broken order},\ }\href@noop {} {\bibfield  {journal} {\bibinfo  {journal} {arXiv preprint arXiv:2505.21614}\ } (\bibinfo {year} {2025})}\BibitemShut {NoStop}%
\bibitem [{\citenamefont {Di~Battista}\ \emph {et~al.}(2024)\citenamefont {Di~Battista}, \citenamefont {Fong}, \citenamefont {D{\'\i}ez-Carl{\'o}n}, \citenamefont {Watanabe}, \citenamefont {Taniguchi},\ and\ \citenamefont {Efetov}}]{di2024infrared}%
  \BibitemOpen
  \bibfield  {author} {\bibinfo {author} {\bibfnamefont {G.}~\bibnamefont {Di~Battista}}, \bibinfo {author} {\bibfnamefont {K.~C.}\ \bibnamefont {Fong}}, \bibinfo {author} {\bibfnamefont {A.}~\bibnamefont {D{\'\i}ez-Carl{\'o}n}}, \bibinfo {author} {\bibfnamefont {K.}~\bibnamefont {Watanabe}}, \bibinfo {author} {\bibfnamefont {T.}~\bibnamefont {Taniguchi}},\ and\ \bibinfo {author} {\bibfnamefont {D.~K.}\ \bibnamefont {Efetov}},\ }\bibfield  {title} {\bibinfo {title} {Infrared single-photon detection with superconducting magic-angle twisted bilayer graphene},\ }\href@noop {} {\bibfield  {journal} {\bibinfo  {journal} {Science Advances}\ }\textbf {\bibinfo {volume} {10}},\ \bibinfo {pages} {eadp3725} (\bibinfo {year} {2024})}\BibitemShut {NoStop}%
\bibitem [{\citenamefont {Jin}\ \emph {et~al.}(2025)\citenamefont {Jin}, \citenamefont {Serpico}, \citenamefont {Lee}, \citenamefont {Confalone}, \citenamefont {Saggau}, \citenamefont {Lo~Sardo}, \citenamefont {Gu}, \citenamefont {Goodge}, \citenamefont {Lesne}, \citenamefont {Montemurro} \emph {et~al.}}]{jin2025exploring}%
  \BibitemOpen
  \bibfield  {author} {\bibinfo {author} {\bibfnamefont {H.}~\bibnamefont {Jin}}, \bibinfo {author} {\bibfnamefont {G.}~\bibnamefont {Serpico}}, \bibinfo {author} {\bibfnamefont {Y.}~\bibnamefont {Lee}}, \bibinfo {author} {\bibfnamefont {T.}~\bibnamefont {Confalone}}, \bibinfo {author} {\bibfnamefont {C.~N.}\ \bibnamefont {Saggau}}, \bibinfo {author} {\bibfnamefont {F.}~\bibnamefont {Lo~Sardo}}, \bibinfo {author} {\bibfnamefont {G.}~\bibnamefont {Gu}}, \bibinfo {author} {\bibfnamefont {B.~H.}\ \bibnamefont {Goodge}}, \bibinfo {author} {\bibfnamefont {E.}~\bibnamefont {Lesne}}, \bibinfo {author} {\bibfnamefont {D.}~\bibnamefont {Montemurro}}, \emph {et~al.},\ }\bibfield  {title} {\bibinfo {title} {Exploring van der waals cuprate superconductors using a hybrid microwave circuit},\ }\href@noop {} {\bibfield  {journal} {\bibinfo  {journal} {Nano Letters}\ }\textbf {\bibinfo {volume} {25}},\ \bibinfo {pages} {3191} (\bibinfo {year} {2025})}\BibitemShut {NoStop}%
\bibitem [{\citenamefont {Thiemann}\ \emph {et~al.}(2018)\citenamefont {Thiemann}, \citenamefont {Beutel}, \citenamefont {Dressel}, \citenamefont {Lee-Hone}, \citenamefont {Broun}, \citenamefont {Fillis-Tsirakis}, \citenamefont {Boschker}, \citenamefont {Mannhart},\ and\ \citenamefont {Scheffler}}]{thiemann2018single}%
  \BibitemOpen
  \bibfield  {author} {\bibinfo {author} {\bibfnamefont {M.}~\bibnamefont {Thiemann}}, \bibinfo {author} {\bibfnamefont {M.~H.}\ \bibnamefont {Beutel}}, \bibinfo {author} {\bibfnamefont {M.}~\bibnamefont {Dressel}}, \bibinfo {author} {\bibfnamefont {N.~R.}\ \bibnamefont {Lee-Hone}}, \bibinfo {author} {\bibfnamefont {D.~M.}\ \bibnamefont {Broun}}, \bibinfo {author} {\bibfnamefont {E.}~\bibnamefont {Fillis-Tsirakis}}, \bibinfo {author} {\bibfnamefont {H.}~\bibnamefont {Boschker}}, \bibinfo {author} {\bibfnamefont {J.}~\bibnamefont {Mannhart}},\ and\ \bibinfo {author} {\bibfnamefont {M.}~\bibnamefont {Scheffler}},\ }\bibfield  {title} {\bibinfo {title} {Single-gap superconductivity and dome of superfluid density in nb-doped srti o 3},\ }\href@noop {} {\bibfield  {journal} {\bibinfo  {journal} {Physical Review Letters}\ }\textbf {\bibinfo {volume} {120}},\ \bibinfo {pages} {237002} (\bibinfo {year} {2018})}\BibitemShut {NoStop}%
\end{thebibliography}%
\end{document}